\documentclass[aps,pra,twocolumn,eqsecnum,groupedaddress,showpacs]{revtex4}

\usepackage{dcolumn}
\usepackage{graphicx}
\usepackage{epsf}
\usepackage{amsmath}
\usepackage{bm}
\usepackage{times}

\newcolumntype{.}{D{x}{}{-1}}
\newcommand{\empt}{\multicolumn{1}{c}{\mbox{---}}}

\def\half{{\textstyle{1\over 2}}}

\def\half{{\textstyle{\frac12}}}
\def\LZa{{\ln\left[\half(Z\alpha)^{-2}\right]}}
\def\LZasquared{{\ln\left[\half(Z\alpha)^{-2}\right]}}

\newcommand{\qsla}{q\hspace{-0.45em}/}

\begin{document}
\preprint{Version 1.0}

\title{Nonrelativistic QED approach to the Lamb shift}

\author{Ulrich D. Jentschura}
\affiliation{Max--Planck--Institut f\"ur Kernphysik,
Saupfercheckweg 1, 69117 Heidelberg, Germany}

\author{Andrzej Czarnecki}
\affiliation{Department of Physics, University of Alberta,
  Edmonton, AB, Canada T6G 2J1}

\author{Krzysztof Pachucki}
\affiliation{Institute of Theoretical Physics,
Warsaw University, Ho\.{z}a 69, 00--681 Warsaw, Poland}

\begin{abstract}
We calculate the one- and two-loop corrections
of order $\alpha\,(Z\,\alpha)^6$ and $\alpha^2\,(Z\,\alpha)^6$ respectively,
to the Lamb shift in hydrogen-like systems using 
the formalism of nonrelativistic quantum electrodynamics. 
We obtain general results valid for all hydrogenic states with nonvanishing 
orbital angular momentum and for the normalized difference of $S$-states.
These results involve the expectation value of local effective operators and 
relativistic corrections to Bethe logarithms.  
The one-loop correction is in agreement with
previous calculations for the particular cases of $S$, $P$, and $D$ states.
The two-loop correction in the order $\alpha^2\,(Z\,\alpha)^6$
includes the pure two-loop self-energy and all diagrams with 
closed fermion loops. The obtained results allow one to obtain improved 
theoretical predictions for all excited hydrogenic states.
\end{abstract}


\maketitle

% \tableofcontents

%
% Introduction
%
\section{Introduction}

The precise calculation of the electron self-energy contribution to energy
levels of hydrogen-like systems is a long-standing problem in 
bound-state quantum electrodynamics. 
The widely used direct numerical 
approach~\cite{Mo1974a,Mo1974b,YeInSh2005} 
is based on a partial-wave decomposition of the 
Dirac-Coulomb propagator, which corresponds to the 
exact all-order treatment of the electron-nucleus interaction.
The one-loop corrections have already been calculated 
to a high numerical precision for a wide range of 
nuclear charge numbers $Z$ (including the case of 
atomic hydrogen $Z=1$), whereas the two-loop correction
has been obtained only for $Z \geq 10$ with limited numerical precision.
The analytic method is based on an expansion in powers
of $Z\,\alpha$ and a subsequent analytic or semianalytic integration.
The two approaches are complementary. In practice,
the numerical method has
primarily been used for systems with a high nuclear charge number,
whereas the analytic method usually provides more accurate predictions 
for low-$Z$ systems. 

Here, we present a unified analytic derivation of
the one- and two-loop binding corrections of order 
$(\alpha/\pi)\,(Z\,\alpha)^6 m \,c^2$ and 
$(\alpha/\pi)^2 \, (Z\,\alpha)^6 m\,c^2$
respectively, for arbitrary bound states of hydrogen-like system
using the formalism of dimensionally regularized
nonrelativistic quantum electrodynamics (NRQED).
This method allows for a natural separation of different energy scales,
(i) the electron mass and (ii) the binding energy, 
using only one regularization 
parameter: the dimension $d$ of the coordinate space. 
This leads to a straightforward
derivation of radiative corrections in terms of expectation values
of some effective operators and the Bethe logarithms.
The calculation  of these operators is the main task of this work,
and we obtain them from standard electromagnetic form factors and the
low-energy limit of the two-photon exchange scattering amplitude. 

This paper is organized as follows: 
In Sec.~\ref{dimnrqed}, dimensionally regularized NRQED 
is outlined. In Sec. III, the one-loop self-energy is derived
by splitting the calculation into low-
(Sec.~\ref{calc1LLEP}), 
middle- (Sec.~\ref{calc1LMEP}), and high-energy parts.
The general one-loop result is presented in Sec.~\ref{calc1LRES},
and the evaluation for $D$, $P$ and $S$ states in 
Secs.~\ref{calc1LnD},~\ref{calc1LnP} and~\ref{calc1LnS},
respectively. The two-loop correction is separated into 
four different gauge-invariant sets of diagrams,
see Figs.~\ref{fig1}---\ref{fig4} below.
These are subsequently investigated in Secs.~\ref{calci}---\ref{calciv}.
Results are summarized in Sec.~\ref{conclusions}.
Moreover, in Appendix~C we present the calculation of an
additional two-loop logarithmic contribution to the ground state
which was omitted in the previous work \cite{Pa2001}.

%
% Dimensionally regularized NRQED
%
\section{Dimensionally regularized NRQED}
\label{dimnrqed}

As is customary in dimensionally regularized QED, we 
assume that the dimension of the space-time 
is $D= 4-2\,\varepsilon$, and that of space 
$d=3-2\,\varepsilon$. The parameter $\varepsilon$ is considered as small,
but only on the level of matrix elements, where an analytic 
continuation to a noninteger spatial dimension is allowed.
Let us briefly discuss the extension of the basic formulas of NRQED to
the case of an arbitrary number of dimensions.
Some basis of dimensionally regularized NRQED in the context
of hydrogen Lamb shift has already been formulated in
\cite{PiSo1998}, however our approach presented below differs in many details.

The momentum-space representation of the photon propagator preserves
its form, namely $g_{\mu\nu}/k^2$. The Coulomb interaction is
\begin{align}
V(r) &= -Z\,e^2\,\int\frac{d^d k}{(2\,\pi)^d}\,
\frac{e^{{\rm i}\,\vec k\cdot \vec r}}{k^2} 
\nonumber\\
&= -\frac{Z\,e^2}{4\,\pi\,r^{1-2\,\varepsilon}}\,
\left[(4\,\pi)^\varepsilon\,
\frac{\Gamma(1-2\,\varepsilon)}{\Gamma(1-\varepsilon)}\right]
\equiv -\frac{Z_\varepsilon\,\alpha}{r^{1-2\,\varepsilon}}\,,
\label{04}
\end{align}
where the latter representation provides an implicit definition of 
$Z_\varepsilon$, and we have
used the formula for the surface area of a $d$-dimensional unit sphere
\begin{equation}
\Omega_d = \frac{2\,\pi^{d/2}}{\Gamma(d/2)}\,.
\end{equation}
The nonrelativistic Hamiltonian of the hydrogenic system is
\begin{equation}
H = \frac{\vec{p}^{\;2}}{2\,m}
-\frac{Z_\varepsilon\,\alpha}{r^{1-2\,\varepsilon}}\,.
\label{06}
\end{equation}
We now turn to relativistic corrections to the Schr\"odinger
Hamiltonian in an arbitrary number of dimensions.
These corrections can be obtained from the Dirac Hamiltonian
by the Foldy-Wouthuysen transformation.
In order to incorporate a part of the radiative effects right from the
beginning, we use an effective Dirac Hamiltonian modified by the
electromagnetic form factors $F_1$ and $F_2$
(see, e.g., Chap.~7 of~\cite{ItZu1980}),
 \begin{align}
\label{HDirac}
H_D =& \vec{\alpha} \cdot
\left[\vec{p} - e \, F_1(\vec \nabla^2) \, \vec{A}\right] + \beta\,m 
+ e\,F_1(\vec \nabla^2) \, A_0
\nonumber\\
& + F_2(\vec \nabla^2) \, \frac{e}{2\,m} \, \left({\rm i}\,\vec{\gamma} \cdot
\vec{E} - \frac{\beta}{2} \, \Sigma^{ij}\,B^{ij} \right)\,,
\end{align}
where
\begin{eqnarray}
B^{ij} &=& \nabla^i\,A^j-\nabla^j\,A^i\,,\label{11}\\
\nabla^i &\equiv& \nabla_i = \partial/\partial x^i\,, \\
\Sigma^{ij} &=& \frac{\rm i}{2}\,[\gamma^i,\gamma^j]\,.
\label{12}
\end{eqnarray}
Formulas for the electromagnetic
form factors $F_{1,2}$ can be found in Appendix~\ref{appa}.
Having the Foldy-Wouthuysen transformation defined by the operator $S$
(see Ref.~\cite{Pa2004}), 
\begin{align}
S =& -\frac{\rm i}{2\,m}\,\left\{
\beta\,\vec\alpha\cdot\vec\pi
-\frac{1}{3\,m^2}\,\beta\,(\vec\alpha\cdot\vec\pi)^3\right.
\nonumber\\
& \left. +\frac{e(1+\kappa)}{2\,m}\,{\rm i}\,\vec\alpha\cdot\vec E
-\frac{e\,\kappa}{8\,m^2}\,
[\vec\alpha\cdot\vec\pi, \beta\,\Sigma^{ij}\,B^{ij}]\right\}\,,
\label{13}
\end{align}
where  $\kappa \equiv F_2(0)$, the new Hamiltonian is obtained via 
\begin{subequations}
\label{trafo}
\begin{equation}
\label{trafo1}
H_{FW} = e^{{\rm i}\,S}\,(H_D-i\,\partial_t)\,e^{-{\rm i}\,S}
\end{equation}
and takes the form
\begin{align}
\label{trafo2}
& H_{FW} = \frac{\vec \pi^{\;2}}{2\,m} + 
e\,[1 + F'_1(0)\,{\vec \nabla}^2] A^0 -
\frac{e}{4\,m}\,(1+\kappa)\,\sigma^{ij}\,B^{ij}
\nonumber\\
& -\frac{\vec \pi^{\;4}}{8\,m^3}
-\frac{e}{8\,m^2}\,(1+2\,\kappa)\,
\left[\vec\nabla\cdot\vec E+\sigma^{ij}\,\{E^i,\pi^j\}\right]
\nonumber \\ 
& -\frac{e}{8\,m^4}\,[F'_1(0)+2\,F'_2(0)]\,\vec\nabla^2\,
\left[\vec\nabla\cdot\vec E+\sigma^{ij}\,\{E^i,\pi^j\}\right]\nonumber \\
&+\frac{\vec{p}^{\,6}}{16\,m^5} +\frac{3+4\,\kappa}{64\,m^4}\,
\bigl\{\vec p^{\;2},\vec\nabla\cdot\vec E+\sigma^{ij}\,\{E^i,\pi^j\}\bigr\}  
\nonumber \\ 
& +\frac{4\,\kappa\,(1+\kappa)-1}{32\,m^3}\,e^2\,\vec E^2
+\ldots
\end{align}
\end{subequations}
The ellipsis denotes the omitted higher-order terms.
We adopt the following
conventions: $\{X,Y\} \equiv X\,Y+Y\,X$, $\vec \pi = \vec p-e\,\vec A$,
$\sigma^{ij} = [\sigma^i,\,\sigma^j]/(2\,{\rm i})$,
and the form factors $F_1$, $F_2$ 
are defined in Eq.~(\ref{A1}) below.
In $d=3$ spatial dimensions, the matrices $\sigma^{ij}$ are equal to 
$\epsilon^{ijk} \, \sigma^k$. The electromagnetic field in $H_{FW}$
is the sum of the external Coulomb field 
and a slowly varying field of the radiation.

There is  an additional correction that cannot be accounted for by
the $F_1$ and $F_2$ form factors. It is represented
by an effective local operator
that is quadratic in the field strengths. This operator is derived separately
by evaluating a low-energy limit of the electron scattering amplitude off
the Coulomb field. An outline of
this calculation are presented in Appendix~\ref{appb}. The result is
\begin{equation}
\label{deltaH}
\delta H =
\frac{e^2}{m^3}\,\vec E^{\;2}\,\chi\,,
\end{equation}
where $E$ is an electric field, and the functions
$\chi \equiv \chi^{(1)} + \chi^{(2)}$ are
given by Eq.~(\ref{eta1plus2}). 

%
% ONE-LOOP ELECTRON SELF-ENERGY
%
\section{ONE--LOOP ELECTRON SELF--ENERGY}
\label{calc1L}

%
% Brief outline of the calculation
%
\subsection{Brief outline of the calculation}
\label{calc1LOUTLINE}

The one-loop electron self-energy contribution in hydrogenlike atoms is
\begin{align}
\delta^{(1)} E = &
\frac{e^2}{\rm i} \,
\int \frac{d^D k}{(2\pi)^D}\,\frac{1}{k^2}
\nonumber\\
&
\times \left< \bar\psi \left|\gamma^\mu\,
\frac{1}{\not\!p-\not\!k-m-\gamma^0\,V}\,\gamma_\mu
\right|\psi\right>
\nonumber\\
& -\delta m\,\langle\bar\psi|\psi\rangle\,.
\label{01}
\end{align}
Here, $p^0 = E_\psi$ is the Dirac energy 
of the reference state, $V$ is the Coulomb potential in $d$ dimensions,
and we use natural 
relativistic units with $\hbar = c = \epsilon_0 = 1$, so that 
$e^2 = 4\pi\alpha$. The electron mass is denoted by $m$,
and $\delta m$ is the one-loop mass counter term.
By $\psi$ we denote the Dirac wave function.
There are three energy scales in Eq. (\ref{01}), 
which imply a natural
separation of the one-loop $\delta^{(1)} E$ into three parts,
\begin{equation}
\label{delta1LEsep}
\delta^{(1)} E = E_L  + E_M  + E_H\,.
\end{equation}
Each part is regularized separately using the same dimensional regularization.
$E_L$ is the low energy part, where the 
photon momentum is of order $ k\sim (Z\,\alpha)^2 \,m$.
$E_M$ is the middle-energy part, where $ k\sim m $,
and the electron momentum is
$p\sim (Z\,\alpha) \,m$. Finally, $E_H$ is a high-energy part where all
loop momenta are of the order of the electron mass.
It is given by the forward three-Coulomb scattering amplitude
and is represented as a local interaction, proportional to $\delta^d(r)$. 

The naming convention for the high-, middle-, and low-energy parts
is a little different from our previous convention.  
E.g., in ~\cite{Pa1993}, the 
contribution referred to as the ``high-energy part'' in this reference 
would correspond to the sum of the ``high-energy part''
and the ``middle-energy part'' in the context of the 
current evaluation. The renaming of the contributions is
influenced by the NRQED-approach used here and by the correspondence
of the different parts to specific effective operators.
In this work, for all operators $Q$,
we consider only the expectation values for states with 
\begin{subequations}
\label{assumptions}
\begin{equation}
 l\neq 0 , 
\end{equation}
and the normalized difference of expectation values 
\begin{equation}
\left< \! \left< Q \right> \! \right> \equiv
n^3 \, \left< nS | Q | nS \right> -
\left< 1S | Q | 1S \right>
\end{equation}
\end{subequations}
for $S$ states. For this reason
the high-energy part $E_H$ vanishes here. Consequently
the ``middle-energy part'' as considered in the current investigation 
corresponds exactly to the ``high-energy part''
of Refs.~\cite{JePa1996,JeSoMo1997}.

The one-loop
bound-state self-energy, for the states under consideration
can be written as
\begin{equation}
\label{struc1L}
\delta^{(1)}E \! = \! \frac{\alpha}{\pi} 
\frac{(Z\alpha)^4}{n^3} \left\{ A_{40} + (Z \alpha)^2 
\left[A_{61}\ln[(Z \alpha)^{-2}] + A_{60} \right]\right\},
\end{equation}
where the indices of the coefficients indicate the 
power of $Z\,\alpha$ and the power of the logarithm,
respectively. The coefficient $A_{40}$ is well known
(for reviews see e.g.~\cite{EiGrSh2001,MoTa2005}),
and we focus here on derivation of the general expression for the 
$\alpha\,(Z\,\alpha)^6$ term.
%
% Low-energy part
%
\subsection{Low-energy part}
\label{calc1LLEP}

In the low energy part, all electron momenta are of the order of
$Z\,\alpha$, so in principle, one could perform a direct 
nonrelativistic expansion of the matrix element
\begin{equation}
\left< \bar\psi \left|\gamma^\mu\,
\frac{1}{\not\!p-\not\!k-m-\gamma^0\,V}\,\gamma_\mu
\right|\psi\right>
\end{equation}
that enters into Eq.~(\ref{01}). It is more convenient however, 
instead of using Eq. (\ref{01}), to take the Dirac 
Hamiltonian with an electromagnetic field and to perform this expansion
by applying the Foldy-Wouthuysen transformation. 
The resulting Hamiltonian, in $d$ dimensions, is given in 
Eq.~(\ref{trafo}).
Here, we can neglect form factors and $H_{FW}$ becomes 
(from now on we will set the electron mass $m$ equal to unity)
\begin{align}
\label{HFW}
& H_{FW} = \frac{\vec\pi^{\,2}}{2}+V(r)
- \frac{e}{4}\,\sigma^{ij}\,B^{ij} - \frac{\vec\pi^{\,4}}{8} +
\frac{\pi}{2}\,Z\,\alpha\,\delta^d(r)
\nonumber \\ 
& +\frac{1}{4}\,\sigma^{ij}\,\nabla^i V\,\pi^j
-\frac{e}{8}\,\Bigl[\vec{\nabla}\cdot\vec E +
\sigma^{ij}\,(E^i\,\pi^j+\pi^j\,E^i)\Bigr]\,. 
\end{align}
The contribution from the Coulomb potential $V$ is explicitly separated 
from the additional electromagnetic fields $\vec E$ and $\vec B$.
The Hamiltonian in Eq. (\ref{HFW}) may be used 
to derive the low-energy part which receives
a natural interpretation as the sum of various relativistic
corrections to the Bethe logarithm. We use the Coulomb gauge for the photon
propagator, and only the transverse part will contribute. This 
treatment of the low-energy part is similar to previous 
calculations~\cite{Pa1993,JePa1996}, 
the difference lies in the presence of dimensional regularization.

The leading nonrelativistic (dipole) low-energy contribution is
\begin{eqnarray}
E_{L0} &=& e^2\int \frac{d^d k}{(2\,\pi)^d\,2\,k}\,
\left(\delta^{ij}-\frac{k^i\,k^j}{k^2}\right)\,
\nonumber \\ &&\times
\left< \phi \left|p^i\,\frac{1}{E-H-k}\,p^j \right|\phi\right>\,,
\label{05}
\end{eqnarray}
where by $H$ we denote the nonrelativistic Hamiltonian in $d$
dimensions, Eq. (\ref{06}).
The wave function $\phi$, in contrast to $\psi$ [see Eq.~(\ref{01})], denotes
the nonrelativistic Schr\"{o}dinger--Pauli wave function.
In the following, we will denote the expectation value
of an arbitrary operator $Q$, 
evaluated with the nonrelativistic Schr\"{o}dinger--Pauli 
wave function, by the shorthand notation $\langle Q \rangle$.

After the $d$-dimensional
integration with respect to $k$, and the expansion in 
$\varepsilon$, $E_{L0}$ becomes \cite{PiSo1998}
\begin{eqnarray}
\label{EL0intermediate}
E_{L0} &=& (4\,\pi)^\varepsilon\,\Gamma(1+\varepsilon)\,
\frac{2\,\alpha}{3\,\pi}\,
 \\ &&\times
\left< \vec p\,(H-E)\left\{\frac{1}{2\,\varepsilon}+\frac{5}{6}-
\ln\left[2(H-E)\right]
\right\}\vec p\,\right> \,,\nonumber
\end{eqnarray}
where we ignore terms of order $\varepsilon$ and higher.
Because the factor $(4\,\pi)^\varepsilon\,\Gamma(1+\varepsilon)$
appears in all the terms, we will drop it out 
consistently in the low-, middle- and high-energy parts,
and as well as in the form factors.
Moreover, in the two-loop calculations discussed below,
we will drop the square of this factor.
The contribution $E_{L0}$ can be rewritten as
\begin{eqnarray}
\label{EL0}
E_{L0} &=&
\frac{4\,\alpha}{3}\, Z\,\alpha\,
\left\{ \frac{1}{2\,\varepsilon} + \frac{5}{6} + \ln[(Z\,\alpha)^{-2}] 
\right\} \, \langle \delta^d(r) \rangle
\nonumber \\ &&
- \frac{4\alpha}{3\pi} \,\frac{(Z\,\alpha)^4}{n^3}\,\ln k_0\,,
\end{eqnarray}
where the second term in this equation involves the Bethe logarithm $\ln k_0$
defined as
\begin{equation}
\frac{(Z\,\alpha)^4}{n^3}\,\ln k_0 = \frac{1}{2}\,\left< \vec p\,(H-E) \,
\ln\left[\frac{2(H-E)}{(Z\,\alpha)^2}\right] \,
\vec p\,\right>\,.
\end{equation}

We consider now all possible relativistic corrections to Eq.~(\ref{EL0}),
and introduce the notation
\begin{align}
\label{deltaQ}
&\delta_Q\,\left< p^i\,\frac{1}{E-H-k}\,p^j \right> \equiv
\biggl< p^i\,\frac{1}{E-H-k}\,(Q-\langle Q\rangle)\,
\nonumber \\ &
\frac{1}{E-H-k}\,p^j \biggr> +
2\, \left< Q\,\frac{1}{(E-H)'}\,p^i\,\frac{1}{E-H-k}\,p^j \right>\,,
\end{align}
where $Q$ is an arbitrary operator. $\delta_Q$ involves the 
first-order perturbations  
to the Hamiltonian, to the energy, and to the wave function.
The first correction $E_{L1}$ is the modification of $E_{L0}$ 
by the relativistic correction to the Hamiltonian,
\begin{equation}
\label{defHR}
H_R = - \frac{\vec p^{\,4}}{8} +
\frac{\pi}{2}\,Z\,\alpha\,\delta^d(r)+\frac{1}{4}\,
\sigma^{ij}\,\nabla^i V\,p^j\,,
\end{equation}
where $\delta^d(r)$ is a $d$-dimensional Dirac delta function.
One could obtain $E_{L1}$ by including this $H_R$ in Eq. (\ref{EL0}).
However, for the comparison with former calculations and for
convenience we will return to Eq. (\ref{05}), 
and split $E_{L1}$ by introducing an 
intermediate cutoff $\Lambda$
\begin{eqnarray}
E_{L1} &=& e^2\left(\int_0^\Lambda + \int_\Lambda^\infty\right)
 \frac{d^d k}{(2\,\pi)^d\,2\,k}\,
\left(\delta^{ij}-\frac{k^i\,k^j}{k^2}\right)\,
\nonumber \\ && \times
\delta _{H_R}\left< p^i\,\frac{1}{E-H-k}\,p^j \right>\,.
\end{eqnarray}
After the $Z\,\alpha$ expansion with $\Lambda = \lambda\,(Z\,\alpha)^2$, 
one goes subsequently to the limits  $\varepsilon\rightarrow 0$ 
and $\lambda\rightarrow \infty$.
Under the assumptions (\ref{assumptions}),
we may perform an expansion in $1/k$ in the second part and obtain
\begin{align}
& E_{L1} = 
\frac{2\,\alpha}{3\,\pi}\,\int_0^\Lambda dk\,k\,
\delta_{H_R}\left< \,\vec p\,\frac{1}{E-H-k}\,\vec p\,\right> 
\\ 
& + \frac{\alpha}{3\,\pi}\,
\left[1+\varepsilon\,\left(\frac{5}{3}-2\,\ln2\right)\right]\,
\int_\Lambda^\infty
dk\,\frac{1}{k^{1+2\,\varepsilon}}\,
\nonumber \\ 
& \left\{\left< [\,\vec p,[\,H_R,\vec p\,]]\right> +
2\,\left< H_R\,\frac{1}{(E-H)'}\,
[\,\vec p,[\,H,\vec p\,]]\right> \right\}\nonumber\,.
\end{align}
After performing the $k$-integration and with the help
of commutator relations it reads
\begin{eqnarray}
\label{EL1intermediate}
E_{L1} &=& 
\frac{\alpha}{\pi}\,\frac{(Z\,\alpha)^6}{n^3}\,\beta_1 +
\frac{\alpha}{3\,\pi}\,
\left\{\frac{1}{2\,\varepsilon}+\frac{5}{6}+
\ln\left[\half\,(Z\,\alpha)^{-2}\right]\right\}
\nonumber \\ 
&& \times \biggl\{\left< \frac{1}{8}\,\vec{\nabla}^4 V +
\frac{\rm i}{4}\,\sigma^{ij}\,p^i
\vec{\nabla}^2 V\,p^j \right> 
\nonumber \\ &&
+2\,\left< H_R\,\frac{1}{(E-H)'}\,\vec{\nabla}^2 V \right> \biggr\}\,.
\end{eqnarray}
Here, $\beta_1$ is
a dimensionless quantity, defined as a finite part of the $k$-integral 
with divergent terms proportional to $\lambda^n$ 
($n = 1, 2, \dots$) and  $\ln(\lambda)$ dropped out in the
limit of large $\lambda$,
\begin{eqnarray}
\label{defbeta1}
\frac{\alpha}{\pi}\,\frac{(Z\,\alpha)^6}{n^3}\,\beta_1 &=& 
\lim_{\lambda\rightarrow\infty}
\frac{2\,\alpha}{3\,\pi}\,
\int_0^\Lambda dk\,k \, 
\nonumber \\ && \times
\delta_{H_R}\,\left< p^i\,\frac{1}{E-H-k}\,p^i \right>\,,
\end{eqnarray}
We recall the relation $\Lambda = \lambda\,(Z\,\alpha)^2$.
In all integrals with an upper limit $\Lambda$ 
to be discussed in the following, 
the divergent terms in $\lambda$ will be subtracted.
Following earlier treatments (e.g.,~\cite{JeEtAl2003}),
we subtract exactly the term proportional
to $\ln(\lambda)$, but not $\ln(2\,\lambda)$.
The presence of the factor
$\half$ under the logarithm in Eq.~(\ref{EL1intermediate})
is a consequence of this subtraction.

The quantity $\beta_1$ can only be calculated numerically.
In constitutes one of three contributions to relativistic Bethe
logarithm $\cal L$, being defined as in~\cite{JeEtAl2003}.
\begin{equation}
\label{defcalL}
{\cal L} = \beta_1 + \beta_2 + \beta_3\,.
\end{equation}
Two others $\beta_2$, $\beta_3$ are defined in Eqs.~(\ref{defbeta2})
and~(\ref{defbeta3}) below.
In this sense, the definition 
of $\beta_1$ in Eq.~(\ref{defbeta1}) corresponds to the definition of 
the low energy part ${\cal L}$ in Eq.~(9) of Ref.~\cite{JeEtAl2003}. 

The second relativistic correction $E_{L2}$ is the 
nonrelativistic quadrupole contribution in the 
conventions adopted in~\cite{Pa1993,JePa1996}.
Specifically, it is the quadratic (in $k$) term
from the expansion of $\exp({\rm i}\,\vec k\cdot\vec r)$,
\begin{eqnarray}
E_{L2} &=& e^2\int \frac{d^d k}{(2\,\pi)^d\,2\,k}\,
\left(\delta^{ij}-\frac{k^i\,k^j}{k^2}\right)\,
\nonumber \\ && \times
\biggl[ \left< p^i\,({\rm i}\,\vec k\cdot\vec r)\,
\frac{1}{E-H-k}\,p^j
\,(-{\rm i}\,\vec k\cdot\vec r)\right>
\nonumber \\ &&
+\left< p^i\,({\rm i}\,\vec k\cdot\vec r)^2\,
\frac{1}{E-H-k}\,p^j\right> \biggr]\,.
\end{eqnarray}
In a similar way as for $E_{L1}$, we split the integration into two parts,
by introducing a cutoff $\Lambda$. In the first part, with the 
$k$-integral from $0$ to $\Lambda$, one can set $d=3$ and extract the 
logarithmic divergence.
In the second part, with the $k$-integral from $\Lambda$ to $\infty$,
we perform a $1/k$ expansion and employ commutator relations,
with the intent of moving the operator $H-E$ to the far left
or right where it vanishes when acting on the Schr\"{o}dinger--Pauli
wave function. In this way we obtain
\begin{widetext}
\begin{eqnarray}
\label{El2intermediate}
E_{L2} &=& 
\frac{\alpha}{\pi}\,\frac{(Z\,\alpha)^6}{n^3}\,\beta_2 +
\frac{\alpha}{\pi}\,\left<
(\vec{\nabla} V)^2\,\frac{2}{3}\,
\left[\frac{1}{\varepsilon}+\frac{103}{60}+
2\, \ln\left[\half\,(Z\,\alpha)^{-2}\right]\right] \right.
\nonumber \\ 
&& \left.
+\vec{\nabla}^4 V\,\frac{1}{40}\,
\left[\frac{1}{\varepsilon}+\frac{12}{5}
+2\, \ln\left[\half\,(Z\,\alpha)^{-2}\right]\right]
+\vec{\nabla}^2V\,\vec p^{\,2}\,\frac{1}{6}\,
\left[\frac{1}{\varepsilon}+\frac{34}{15}
+2\, \ln\left[\half\,(Z\,\alpha)^{-2}\right]\right]
\right> \,.
\end{eqnarray}
Here, $\beta_2$ is defined as the finite part of the integral
[see the discussion following Eq.~(\ref{defbeta1})]
\begin{eqnarray}
\label{defbeta2}
\frac{\alpha}{\pi}\,\frac{(Z\,\alpha)^6}{n^3}\,\beta_2 = 
4\,\pi\,\alpha\,\,\lim_{\lambda\rightarrow\infty} \int_0^\Lambda\frac{d^3k}{(2\,\pi)^3\,2\,k}\,
\left(\delta^{ij}-\frac{k^i\,k^j}{k^2}\right)\,
\left\{\left< \,p^i\,({\rm i}\,\vec k\cdot\vec r)^2\,
\frac{1}{E-H-k}\,p^j\,\right> \right.
\nonumber \\ 
\left. +\left< \,p^i\,({\rm i}\,\vec k\cdot\vec r)\,
\frac{1}{E-H-k}\,p^j\,(-{\rm i}\,\vec k\cdot\vec r)\,\right>
\right\}\,.
\end{eqnarray}

The third contribution $E_{L3}$ originates from the relativistic
corrections to the coupling of the electron to the electromagnetic field.
These corrections can be obtained  from the Hamiltonian in Eq.~(\ref{04}),
and they have the form of a correction to the current
\begin{equation}
\delta j^i = -\frac{1}{2}\,p^i\,\vec p^{\,2}
+\frac{1}{2}\,\sigma^{ij}\,k^j\,\vec k\cdot\vec r
+\frac{\rm i}{4}\,\sigma^{ij}\,k\,p^j
-\frac{1}{4}\,\sigma^{ij}\,\nabla^j V\,.
\end{equation}
The corresponding correction $E_{L3}$ is
\begin{eqnarray}
E_{L3} &=& 2\,e^2\int \frac{d^d k}{(2\,\pi)^d\,2\,k}\,
\left(\delta^{ij}-\frac{k^i\,k^j}{k^2}\right)\,
\left< \delta j^i\frac{1}{E-H-k}\,p^j \right>\,.
\end{eqnarray}
We now perform an angular averaging of the matrix element,
replace $k$ in the numerator by $E-H$, and use commutator 
relations to bring the correction $E_{L3}$ into the form
\begin{eqnarray}
E_{L3} =-2\,e^2\,\frac{d-1}{d}\,\int \frac{d^d k}{(2\,\pi)^d\,2\,k}\,
\left< \left(\frac{p^k\,{\vec p}^{\,2}}{2}+\frac{d-2}{d-1}\,
\frac{\sigma^{kl}\,\nabla^l V}{2}\right)
\frac{1}{E-H-k}\,p^k \right>\,.
\end{eqnarray}
We again split this integral into two parts.  In the first part
$k<\Lambda$, one can approach the limit $d=3$, and
in the second part $k>\Lambda$ one performs a $1/k$-expansion
and obtains
\begin{eqnarray}
\label{EL3intermediate}
E_{L3} &=& 
\frac{\alpha}{\pi}\,\frac{(Z\,\alpha)^6}{n^3}\,\beta_3
-\frac{4\alpha}{3\,\pi}
\,\left[\frac{1}{2\,\varepsilon}+
\frac{5}{6}+
\ln\left[\half\,(Z\,\alpha)^{-2}\right]\right]
\left<\frac{1}{4}\,\vec{\nabla}^2V\,\vec p^{\,2} +
\frac{1}{2}\,\bigl(\vec{\nabla} V\bigr)^2\right>\,,
\end{eqnarray}
where $\beta_3$ is the finite part of the integral
\begin{equation}
\label{defbeta3}
\frac{\alpha}{\pi}\,\frac{(Z\,\alpha)^6}{n^3}\,\beta_3 = 
-\frac{4\,\alpha}{3\,\pi}\,\lim_{\lambda\rightarrow\infty} 
\int_0^\Lambda dk\,k
\left< \left(\frac{1}{2}\,p^i\,p^2+\frac{1}{4}\,\sigma^{ij}\,\nabla^j
V\right)\,\frac{1}{E-H-k}\,p^i\,\right>\,.
\end{equation}
This completes the treatment of the low energy part, which is
\begin{equation}
\label{defEL}
E_L = E_{L1}+E_{L2}+E_{L3}\,.
\end{equation}

%
% Middle-energy part
%
\subsection{Middle-energy part}
\label{calc1LMEP}

We here consider the middle-energy part
$E_M$ as the contribution originating from photon momentum of the order
of the electron mass and electron momenta of order $Z\,\alpha$.
In this momentum region, radiative corrections can be effectively
represented by electron form factors and higher-order structure functions.
Electron form factors $F_1$ and $F_2$ modify the coupling of the Dirac
electron to the electromagnetic field and the 
resulting effective Hamiltonian is given in Eq.~(\ref{HDirac}).
Here we assume that $\vec A=0$, $A^0$ represents
a static Coulomb potential,  and 
$\vec{E} = -\vec{\nabla} A^0$
is the electric field of the nucleus. 
One finds a nonrelativistic expansion by the Foldy-Wouthuysen transformation
in Eq. (\ref{13}), and the resulting Hamiltonian [see 
Eq.~(\ref{trafo2})] after putting $\vec A=0$ and neglecting $F_2(0)^2$ is 
\begin{eqnarray}
H_{FW} &=& \frac{\vec p^{\,2}}{2} + 
e\,F_1(\vec{\nabla}^2)\,A^0 -
\frac{\vec p^{\,4}}{8}
-\frac{e}{8}\,[F_1(\vec{\nabla}^2)+2\,F_2(\vec{\nabla}^2)]\,
\Bigl(\vec{\nabla}\cdot\vec E + 2\,\sigma^{ij}\,E^i\,p^j\Bigr)
\nonumber \\ && +\frac{\vec p^{\,6}}{16}
+\frac{e}{64}\,[3+4\,F_2(0)]
\Bigl\{\vec p^{\,2},\vec{\nabla}\cdot\vec E + 2\,\sigma^{ij}\,E^i\,p^j\Bigr\}
-\frac{1-4\,F_2(0)}{32}\,e^2\,\vec E^2\,.
\label{21}
\end{eqnarray}
The leading $(\alpha/\pi) \, (Z\,\alpha)^4$ one-loop correction reads
\begin{equation}
\label{defEM0}
E_{M0} = \langle\delta^{\rm (1)}V\rangle\,,
\end{equation}
where the ``radiative potential'' $\delta V$ is defined as
\begin{equation}
\label{defdeltaV}
\delta V = \left[F'_1(0) + \frac{1}{4}\,F_2(0)\right]\,
\vec{\nabla}^2 V
+\frac{F_2(0)}{2}\,\sigma^{ij}\,\nabla^i V\,p^j\,,
\end{equation}
and the superscript ${\rm (1)}$ in Eq.~(\ref{defEM0})
denotes the one-loop component of $\delta V$.
The expansion of $F_i$ in powers of $q^2$ is obtained
in Eq.~(\ref{ff}). Using these results, one obtains
\begin{equation}
\label{EM0}
E_{M0} = 
-\frac{1}{6\,\varepsilon}\,\frac{\alpha}{\pi}\,
\langle \vec{\nabla}^2 V\rangle
+\frac{\alpha}{4\,\pi}\,\langle\sigma^{ij}\nabla^i V\,p^j\rangle\,.
\end{equation}
Together with the low-energy part $E_{L0}$ in Eq. (\ref{EL0}),
this gives
\begin{eqnarray}
\label{defE0}
E_0 \equiv E_{L0}+E_{M0} &=& \frac{\alpha}{\pi}\,(Z\,\alpha)^4 \,
\left[ \frac{10}{9}+\frac{4}{3}\,\ln\left[(Z\,\alpha)^{-2}\right]\right]\,
\frac{\delta_{l0}}{n^3}
+\frac{\alpha}{4\,\pi}\,
\left< \sigma^{ij}\nabla^i V\,p^j \right>
-\frac{4\,\alpha}{3\,\pi}\,\frac{(Z\,\alpha)^4}{n^3}\,\ln k_0\,,
\end{eqnarray}
which is the well-known leading 
$(\alpha/\pi) \, (Z\,\alpha)^4$ contribution to the hydrogen Lamb shift.

Let us now consider the one-loop correction
of relative-order $(Z\,\alpha)^2$. The first contribution
$E_{M1}$ comes from the one-loop form factors $F_1$ and $F_2$ 
in Eq. (\ref{trafo2}) combined with the relativistic correction 
to the wave function:
\begin{eqnarray}
\label{defEM1}
E_{M1} &=& 2\,\left<
\left\{\left[F'^{\rm (1)}_1(0)+\frac{1}{4}\,F^{\rm (1)}_2(0)\right]\,
\vec{\nabla}^2 V 
+ \frac{F_2^{\rm (1)}(0)}{2}\,\sigma^{ij}\,\nabla^i V\,p^j\right\}
\,\frac{1}{(E-H)'}\,H_R\right>
\nonumber \\ &&
+\frac{F'^{\rm (1)}_1(0)+2 F'^{\rm (1)}_2(0)}{8}\,
\langle\vec{\nabla}^4V+2\,{\rm i}\,
\sigma^{ij}\,p^i\vec{\nabla}^2 V\, p^j\rangle
+ F''^{\rm (1)}_1(0)\,\langle\vec{\nabla}^4V \rangle
\nonumber \\ &&
-\frac{F^{\rm (1)}_2(0)}{16}\,\left<
\Bigl\{\vec p^{\,2},\vec{\nabla}^2V
+2\,\sigma^{ij}\,\nabla^i V\,p^j \Bigr\}\right>
+\frac{F^{\rm (1)}_2(0)}{8}\,\langle(\vec{\nabla} V)^2\rangle\,.
\end{eqnarray}
By the superscript (1), we denote the one-loop
component of the form factors, as given in 
Eq.~(\ref{ff2}) in Appendix~\ref{appa}.

The second contribution $E_{M2}$ comes from an additional 
term $\delta^{(1)} H$
in the NRQED Hamiltonian, see Eq. (\ref{deltaH}),
\begin{equation}
\label{defH1L}
\delta^{\rm (1)} H = 
\left(\frac{1}{6} - \frac{1}{3\,\varepsilon} \right)
\,\frac{\alpha}{\pi}\,(\vec{\nabla} V)^2\,.
\end{equation}
The corresponding correction to the energy is
\begin{equation}
\label{defEM2}
E_{M2} = \left< \delta^{\rm (1)} H \right>\,,
\end{equation}
and the total $E_M$ contribution is
\begin{equation}
\label{defEM}
E_M = E_{M1} +  E_{M2}\,.
\end{equation}
%
% One-loop result
%
\subsection{General one-loop result}
\label{calc1LRES}

We may now present the complete
one-loop correction $\delta^{\rm (1)} E$ up to the order
$(\alpha/\pi)\,(Z\,\alpha)^6$. It is a sum of
the low-energy term $E_L$ given in Eq.~(\ref{defEL}), 
the middle-energy term $E_M$ in Eq. (\ref{defEM}), and
the lower-order term $E_0$ as defined in Eq.~(\ref{defE0}),
\begin{eqnarray}
\label{delta1LE}
\delta^{\rm (1)} E &=&
\frac{\alpha}{\pi}\,\frac{(Z\,\alpha)^4}{n^3} \,
\left\{ 
\left[ \frac{10}{9}+\frac{4}{3}\,\ln\left[(Z\,\alpha)^{-2}\right]\right]\,
\delta_{l0} - \frac43\,\ln k_0 \right\}
+\frac{Z\,\alpha^2}{4\,\pi}\,
\left< \sigma^{ij} \nabla^i V p^j \right>
\nonumber \\ 
&& + \frac{\alpha}{\pi}\,\left\{\frac{(Z\,\alpha)^6}{n^3}\, {\cal L}
+\left(\frac{5}{9}+
\frac{2}{3}\,\ln\left[\half\,(Z\,\alpha)^{-2}\right]\right)\,
\left<\vec{\nabla}^2 V\,\frac{1}{(E-H)'}\,H_R\right>
+\frac{1}{2}\,\left< \sigma^{ij}\nabla^i
V p^j\,\frac{1}{(E-H)'}\,H_R\right> \right.
\nonumber \\ 
&& +\left(\frac{779}{14400}+\frac{11}{120}\,
\ln\left[\half\,(Z\,\alpha)^{-2}\right]\right)
\,\langle \vec{\nabla}^4 V\rangle
+\left(\frac{23}{576}+\frac{1}{24}\,
\ln\left[\half\,(Z\,\alpha)^{2}\right]\right)
\,\langle 2 \, {\rm i}\,\sigma^{ij}\,p^i\vec{\nabla}^2V p^j \rangle
\nonumber \\ &&
+ \left. \left(\frac{589}{720}+\frac{2}{3}\,
\ln\left[\half\,(Z\,\alpha)^{2}\right]\right)\,
\langle(\vec{\nabla} V)^2\rangle
+\frac{3}{80}\,\bigl\langle\vec p^{\,2}\, \vec{\nabla}^2 V\bigr\rangle
-\frac{1}{8}\,\bigl\langle \vec{p}^{\,2}\, 
\sigma^{ij}\,\nabla^i V p^j \bigr\rangle\right\}\,.
\end{eqnarray}
\end{widetext}
The first two terms corresponds to the $\alpha\,(Z\,\alpha)^4$ 
term in Eq.~(\ref{struc1L}), whereas the latter terms 
give the $\alpha\,(Z\,\alpha)^6$ contribution.
The relativistic Bethe Logarithm ${\cal L}$, 
defined in Eq.~(\ref{defcalL}), consists of a sum of
$\beta_1$ defined in Eq.~(\ref{defbeta1}), $\beta_2$ in Eq.~(\ref{defbeta2}),
and $\beta_3$  in Eq.~(\ref{defbeta3}).
For the convenience of the reader we briefly
recall that all matrix elements should be evaluated in $d=3$ 
spacetime dimensions, which implies
\begin{subequations}
\begin{align}
\vec{\nabla}^2 V & \to 4\pi Z\,\alpha \, \delta^3(r)\,,\\
\sigma^{ij} \nabla^i V p^j &\to 
Z\,\alpha\, \frac{\vec{\sigma} \cdot \vec{L}}{r^3}\,,\\
\sigma^{ij} p^i \nabla^2 V p^j & \to
  4 \pi  Z\,\alpha \, \vec{p} \times [\delta^3(r) \vec{p}]\,,\\
\sigma^{ij} \nabla^j V & \to Z\,\alpha \frac{\vec{r} \times \vec{\sigma}}{r^3}\,.
\end{align}
\end{subequations}
This concludes the calculation of the one-loop electron self-energy. The
matrix elements entering into (\ref{delta1LE}) are 
evaluated below in Secs.~\ref{calc1LnD}, \ref{calc1LnP}, 
and~\ref{calc1LnS} for a number of hydrogenic 
states and compared to results previously obtained in the literature.

%
% $D$ states
%
\subsection{Results for $D$ states}
\label{calc1LnD}

Our aim is to give a few numerical results for some 
phenomenologically important hydrogenic states,
based on the general result~(\ref{delta1LE}).
For $D$ states, the wave function behaves at the origin as $\sim r^2$. 
This means that a few matrix elements, 
such as $\langle \vec{\nabla}^4 V\rangle$, 
are actually vanishing. The following is a list of 
the nonvanishing matrix elements for $l=2$:
\begin{subequations}
\label{matelemD}
\begin{align}
& \left< nD \left| 
\frac{Z\,\alpha}{r^3}\,\frac{1}{(E-H)'}\,\vec{p}^{\,4}
\right| nD \right>  
\nonumber\\
& \qquad = \frac{(Z\,\alpha)^6}{n^3}\,\left(-\frac{1118}{7875}
-\frac{4}{25\,n}
+\frac{86}{105\,n^2}\right) \,, \\[0.5ex]
& \left< nD \left| 
\frac{Z\,\alpha}{r^3}\,\frac{1}{(E-H)'}\,\frac{Z\,\alpha}{r^3}
\right| nD \right>  
\nonumber\\
& \qquad = \frac{(Z\,\alpha)^6}{n^3}\,\left(-\frac{709}{94500}
-\frac{1}{150\,n}
+\frac{1}{105\,n^2}\right) \,, \\[0.5ex]
& \left< nD \left| 
\frac{(Z\,\alpha)^2}{r^4}
\right| nD \right>  
= \frac{(Z\,\alpha)^6}{n^3}\,\frac{2(n^2-2)}{105\,n^2} \,, \\[0.5ex]
& \left< nD \left| 
\vec{p}^{\,2}\,\frac{Z\,\alpha}{r^3}
\right| nD \right>  
= \frac{(Z\,\alpha)^6}{n^3}\,
\left(\frac{4}{105}-\frac{1}{7\,n^2}\right)\,.
\end{align}
\end{subequations}
All of the above are evaluated on the 
nonrelativistic Schr\"{o}dinger wave function.
They are  finite so that one may set the space dimension equal to three.
The final results for the different fine-structure sublevels are
\begin{subequations}
\label{res1LnD}
\begin{align}
\label{res1LnD32}
& A_{60}(nD_{3/2}) + A_{61}(nD_{3/2}) \ln\left[(Z\alpha)^{-2}\right]
\nonumber\\
& \qquad = {\cal L}(nD_{3/2})
- \frac{157}{30240}-\frac{3}{80 n}+\frac{3007}{37800 n^2}
\nonumber\\
& \qquad \qquad + \frac{4}{315}\,\left(1-\frac{2}{n^2}\right)
\ln\left[\half(Z\,\alpha)^{-2}\right],
\end{align}
and
\begin{align}
\label{res1LnD52}
& A_{60}(nD_{5/2}) + A_{61}(nD_{5/2}) \ln\left[(Z\alpha)^{-2}\right]
\nonumber\\
& \qquad = {\cal L}(nD_{5/2})
+ \frac{379}{18900}+\frac{1}{60 n}-\frac{1759}{18900 n^2}
\nonumber\\
& \qquad \qquad + \frac{4}{315}\,\left(1-\frac{2}{n^2}\right)
\ln\left[\half(Z\alpha)^{-2}\right].
\end{align}
\end{subequations}
They are in agreement with results reported previously in Eqs.~(12c) and~(12d)
of~\cite{JeEtAl2003}.  Values for ${\cal L}(nD_{3/2})$ and 
${\cal L}(nD_{5/2})$ can be found in Table~1 of Ref.~\cite{JeEtAl2003}.

%
% $P$ states
%
\subsection{Results for $P$ states}
\label{calc1LnP}

For $P$ states, a few more of the matrix elements in Eq.~(\ref{delta1LE})
are nonvanishing, and we have
\begin{subequations}
\label{matelemP}
\begin{align}
& \left< nP \left| \frac{Z\,\alpha}{r^3}\,\frac{1}{(E-H)'}\,\vec{p}^{\,4} 
\right| nP \right>
\nonumber\\
& \qquad = \frac{(Z\,\alpha)^6}{n^3}\,
\left(-\frac{346}{135}-\frac{4}{3\,n}+\frac{22}{5\,n^2}\right)\,, \\
& \left< nP \left| \frac{Z\,\alpha}{r^3}\,
\frac{1}{(E-H)'}\,\frac{Z\,\alpha}{r^3}
\right| nP \right> =
\nonumber\\
& \qquad =\frac{(Z\,\alpha)^6}{n^3}\,
\left(-\frac{227}{540}-\frac{1}{6\,n}+\frac{1}{5\,n^2}\right)\,, \\
& \left< nP \left| \frac{(Z\,\alpha)^2}{r^4}
\right| nP \right> = \frac{(Z\,\alpha)^6}{n^3}\,
\frac{2\,(3\,n^2-2)}{15\,n^2}\,, \\
& \left< nP \left| \vec{p}^{\,2}\,\frac{Z\,\alpha}{r^3}
\right| nP \right> 
= \frac{(Z\,\alpha)^6}{n^3}\,
\left(\frac{4}{5}-\frac{13}{15\,n^2}\right) \,, \\
& \left< nP \left| \vec{\nabla}^2\,[4\,\pi\,(Z\,\alpha)\,\delta^3(r)]
\right| nP \right> 
\nonumber\\
& \qquad = 
\frac{(Z\,\alpha)^6}{n^3}\,\frac{8}{3}\,\left(1-\frac{1}{n^2}\right) \,,\\
& \left< nP_J \left| {\rm i}\,\sigma^{ij}\,p^i 
[4\,\pi\,(Z\,\alpha)\,\delta^3(r)]\,
p^j \right| nP_J \right> 
\nonumber\\
& \qquad = \left< nP_J \left| \vec\sigma\cdot\vec L \right| nP_J \right>\,
\frac{4}{3}\,\frac{(Z\,\alpha)^6}{n^3}\,\frac{(1-n^2)}{n^2}\,.
\end{align}
\end{subequations}
The results for the different fine-structure sublevels are
\begin{subequations}
\begin{align}
& A_{60}(P_{1/2}) + A_{61}(P_{1/2}) \ln\left[(Z\,\alpha)^{-2}\right]
\nonumber\\
& \qquad = 
{\cal L}(nP_{1/2})
+\frac{637}{1800}
-\frac{1}{4\,n}
-\frac{767}{5400\,n^2}
\nonumber\\
&  \qquad \qquad +\left(\frac{11}{15} - \frac{29}{45\,n^2}\right)\,
\ln\left[\half\,(Z\,\alpha)^{-2}\right],
\end{align}
and
\begin{align}
& A_{60}(P_{3/2}) + A_{61}(P_{3/2}) \ln\left[(Z\,\alpha)^{-2}\right]
\nonumber\\
& \qquad = {\cal L}(nP_{3/2})
+\frac{2683}{7200}
+\frac{1}{16\,n}
-\frac{2147}{5400\,n^2}
\nonumber\\
& \qquad \qquad +\left(\frac{2}{5}-\frac{14}{45\,n^2}\right)\,
\ln\left[\half\,(Z\,\alpha)^{-2}\right].
\end{align}
\end{subequations}
As for $D$ states, the values for ${\cal L}(nP_{1/2})$ and
${\cal L}(nP_{3/2})$ can then be found in Table~1 of~\cite{JeEtAl2003},
and the polynomials in $n^{-1}$ which are part of the 
above results are consistent with those reported in 
Eqs.~(12a) and~(12b) of Ref.~\cite{JeEtAl2003}.

%
% Normalized difference of $S$ states
%
\subsection{Results for the normalized difference of $S$ states}
\label{calc1LnS}

Considering the following matrix elements for $l=0$ of the 
$S$-state normalized difference $\langle\!\langle \cdot \rangle\!\rangle$,
as defined in Eq.~(\ref{assumptions}), we obtain
\begin{subequations}
\label{matelemS}
\begin{align}
& \left< \!\!\! \left<
4\,\pi\,(Z\,\alpha)\,\delta^3(r)\,\frac{1}{(E-H)'}\,\vec{p}^4
\right> \!\!\! \right>
\nonumber\\
&\quad = 32\,(Z\,\alpha)^6\,
\left[-\frac{1}{4}-\frac{1}{n}+\frac{5}{4\,n^2}+
\gamma+\Psi(n)-\ln n\right]\,,\\
& \left< \!\!\! \left<
4\pi(Z\,\alpha) \delta^3(r)\,\frac{1}{(E-H)'}\,
4\pi(Z\,\alpha)\delta^3(r)
\right> \!\!\! \right>
\nonumber\\
&\quad =
16\,(Z\,\alpha)^6\,
\left[1-\frac{1}{n}+\gamma+\Psi(n)-\ln n\right]\,,
\end{align}
\begin{align}
& \left< \!\!\!\!\: \left< \vec{\nabla}^2[4\pi(Z\,\alpha)\,\delta^3(r)]
\right> \!\!\!\!\: \right> = 8\,(Z\,\alpha)^6\,
\frac{1-n^2}{n^2}\,,\\
& \left< \!\!\! \left<
\frac{(Z\,\alpha)^2}{r^4}
\right> \!\!\! \right> 
\nonumber\\
&\quad =
8(Z\alpha)^6\left[
-\frac{2}{3}+\frac{1}{2n}+\frac{1}{6n^2}+\gamma+\Psi(n)-\ln n\right]\,.
\end{align}
\end{subequations}
Here, $\gamma = 0.577216\dots$ is Euler's constant.
One finally obtains the following result 
for the general normalized difference of the self-energy
for $S$ states,
\begin{eqnarray}
\label{resnS}
\lefteqn{A_{60}(nS) - A_{60}(1S) +
\left[ A_{61}(nS) - A_{61}(1S) \right]
\ln\left[(Z\alpha)^{-2}\right]}
\nonumber\\
& & = {\cal L}(nS_{1/2}) 
- {\cal L}(1S_{1/2}) 
- \frac{16087}{5400}+\frac{263}{60\,n}-\frac{7583}{5400\,n^2}
\nonumber \\
& & + \frac{163}{30}\,[\gamma+\Psi(n)-\ln n]
+ \ln\left[\half (Z\,\alpha)^{-2}\right]
\nonumber \\
& & \times \left\{ 
-\frac{103}{45}+\frac{4}{n}-\frac{77}{45n^2}
+4\left[\gamma+\Psi(n)-\ln n\right]\right\}.
\end{eqnarray}
Here, $\Psi(x) = \Gamma'(x)/\Gamma(x)$
is the logarithmic derivative of the Euler Gamma function.
Values for $A_{60}(nS_{1/2})$ in the range $n = 1,\dots,8$
have been obtained using the above formula (\ref{resnS}) 
and a generalization of methods 
used previously
for states with nonvanishing angular momentum quantum numbers
(see Table~\ref{table1}).
\begin{widetext}
\begin{center}
\begin{table}[htb!]
\begin{minipage}{16.0cm}
\caption{\label{table1} Detailed breakdown of the contributions to 
$A_{60}(nS)$, obtained with the help of Eq.~(\ref{resnS}).
The results for $A_{60}(1S)$ and $A_{60}(2S)$ are obtained here with 
an increased accuracy as compared to Ref.~\cite{Pa1993}.
The generalized Bethe logarithms $\beta_1$, $\beta_2$ and $\beta_3$ are defined 
in Eqs.~(\ref{defbeta1}),~(\ref{defbeta2}) and~(\ref{defbeta3}),
respectively. The contribution ${\cal H}$ is a contribution to $A_{60}$ from 
high-energy virtual photons, given in Eq.~(5.116) of Ref.~\cite{Pa1993} for the 
$1S$ state and generalized to arbitrarily high principal quantum numbers 
using Eq.~(\ref{resnS}). We have $A_{60}(nS) = {\cal L}(nS) + {\cal H}(nS)$ and 
recall that ${\cal L} = \sum_{i=1}^3 \beta_i$.}
\begin{tabular}{rl@{\hspace{0.2in}}l@{\hspace{0.2in}}l@{\hspace{0.2in}}%
l@{\hspace{0.2in}}l@{\hspace{0.2in}}l}
\hline
\hline
n & 
\multicolumn{1}{c}{$\beta_1(nS)$} &
\multicolumn{1}{c}{$\beta_2(nS)$} &
\multicolumn{1}{c}{$\beta_3(nS)$} &
\multicolumn{1}{c}{${\cal L}(nS)$} &
\multicolumn{1}{c}{${\cal H}(nS)$} &
\multicolumn{1}{c}{$A_{60}(nS)$} \\
\hline
1 &  -3.268\,213\,21(1) & -40.647\,026\,69(1) &  16.655\,330\,43(1) & -27.259\,909\,48(1) &  -3.664\,239\,98 & -30.924\,149\,46(1) \\
2 &  -6.057\,407\,04(1) & -39.829\,658\,28(1) &  17.536\,099\,97(1) & -28.350\,965\,35(1) &  -3.489\,499\,74 & -31.840\,465\,09(1) \\
3 &  -6.213\,948(1) &     -39.669\,430(1) &      17.656\,995(1) &   -28.226\,383(1) &      -3.476\,117&     -31.702\,501(1) \\
4 &  -6.167\,093(1) &     -39.611\,903(1) &      17.695\,346(1) &   -28.083\,650(1) &      -3.478\,272&     -31.561\,922(1) \\
5 &  -6.100\,341(1) &     -39.584\,944(1) &      17.712\,334(1) &   -27.972\,951(1) &      -3.482\,442&     -31.455\,393(1) \\
6 &  -6.039\,851(1) &     -39.570\,199(1) &      17.721\,349(1) &   -27.888\,701(1) &      -3.486\,429&     -31.375\,130(1) \\
7 &  -5.988\,793(1) &     -39.561\,272(1) &      17.726\,711(1) &   -27.823\,354(1) &      -3.489\,870&     -31.313\,224(1) \\
8 &  -5.946\,180(1) &     -39.555\,462(1) &      17.730\,161(1) &   -27.771\,481(1) &      -3.492\,776&     -31.264\,257(1) \\
\hline
\hline
\end{tabular}
\end{minipage}
\end{table}
\end{center}
\end{widetext}
One observes the somewhat irregular behavior
of $\beta_1$ as a function of $n$, which is partially 
compensated by the other contributions to 
$A_{60}$. Compared to other 
families of states with the same angular momenta but
varying principal quantum number~\cite{JeEtAl2003},
the $A_{60}$ for $S$ states display a rather 
unusual behavior as a function 
of $n$, with a minimum between $n=2$ and $n=3$.
The calculations of the relativistic 
Bethe logarithms ${\cal L}$, for higher excited $S$ states, 
are quite involved and will be described in detail 
elsewhere. The value for $1S$ as reported 
in Table~\ref{table1} represents an improved 
result (with a numerically small correction)
as compared to the result communicated in Ref.~\cite{Pa1993},
as already detailed in~\cite{JeMoSo1999}.  
For $n \geq 3$, the results for $A_{60}$ have
not appeared in the literature to the best of our knowledge.
The results for $n=3$ and $n=4$ are consistent with numerical 
results for the self-energy remainder function as reported 
in Ref.~\cite{JeMo2004pra} for these states.

%
% TWO-LOOP ELECTRON SELF-ENERGY
%
\section{TWO--LOOP ELECTRON SELF--ENERGY}
\label{calci}

%
% Calculation
%
\subsection{Calculation}
\label{calciCALC}

\begin{figure}[htb]%
\begin{center}\includegraphics[width=0.7\linewidth]{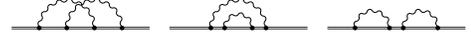}\end{center}%
\caption{\label{fig1}
Pure two-loop self-energy diagrams (subset $i$ of the 
two-loop diagrams). The double line denotes the 
bound-electron propagator.}%
\end{figure}

The two-loop
bound-state energy shift, for the states under investigation here,
can be written as
\begin{align}
\label{struc2L}
& \delta^{\rm (2)}E = \left(\frac{\alpha}{\pi}\right)^2 \,
\frac{(Z\,\alpha)^4}{n^3} \left\{B_{40} \right.
\nonumber\\[2ex]
& \quad \left. +  (Z \alpha)^2 
\left[B_{62}\ln^2[(Z\, \alpha)^{-2}] + 
B_{61}\ln[(Z \alpha)^{-2}] + B_{60} \right]\right\}.
\end{align}
Here, the indices of the coefficients indicate the
power of $Z\,\alpha$ and the power of the logarithm,
respectively. The coefficient $B_{40}$ is well known
(for reviews see e.g.~\cite{EiGrSh2001,MoTa2005}),
and we focus here on general expressions for the $\alpha^2\,(Z\,\alpha)^6$ 
coefficient. 
We split the calculation into four parts, labeled $i$---$iv$ according 
to the subsets of diagrams in Figs.~\ref{fig1}---\ref{fig4}.
This entails a separation of the two-loop energy shift
according to 
\begin{equation}
\delta^{\rm (2)}E = \delta^{\rm (2)}E^i + 
\delta^{\rm (2)}E^{ii} + 
\delta^{\rm (2)}E^{iii} +
\delta^{\rm (2)}E^{iv}\,.
\end{equation}
The specific contributions will be considered subsequently in the 
following sections of this article. The $B$-coefficients
corresponding to the subsets $i$---$iv$ will be distinguished 
using appropriate superscripts.

We first focus on the pure two-loop self-energy diagrams 
as shown in Fig.~\ref{fig1} and denote the corresponding
energy shifts and $B$-coefficients by a superscript $i$.
As compared to the one-loop case treated in Sec.~\ref{calc1L},
the two-loop calculation involves a few more terms with regard to the 
form-factor contributions. However, as it has been stressed in 
Refs.~\cite{PaJe2003,Je2004b60},
the leading order of the two-loop low-energy part 
is already $(\alpha/\pi)^2 \, (Z\alpha)^6$, so there 
are no relativistic or quadrupole corrections to include 
at this energy scale. More precisely,
we split the two-loop contribution into four parts~\cite{Pa2001}:
\begin{equation}
\label{split2L}
\delta^{\rm (2)} E^i = E_L + E_M + E_F + E_H\,.
\end{equation}
Here, the contributions $E_L$, $E_M$ and $E_H$ 
are appropriately redefined 
for the two-loop problem [cf.~Eq.~(\ref{delta1LEsep}) for the 
one-loop case]. We use definitions local to the current Section
for the specific contributions.

The two-loop 
$E_H$ is a high-energy part given by a two-loop forward scattering amplitude
with three Coulomb vertices. Because it leads to 
a local potential (proportional to a Dirac $\delta$ in coordinate space),
the term $E_H$ does not contribute to the energy
of states with $l\neq 0$ or to the normalized difference
of $S$-states. So, we will not consider this contribution here.
For $S$ states, this term gives an $n$-independent contribution to the 
nonlogarithmic term $B_{60}$.

The form-factor contribution $E_F$ corresponds to an integration region 
where both photon momenta are of the order of
the electron mass, but the electron momentum is of the order 
of $Z\,\alpha$. This part is a sum of two terms:
\begin{equation}
\label{defEF}
E_F = E_{F1}+E_{F2}\,.
\end{equation}
The first term $E_{F1}$
comes from two-loop form factors, in the same way as 
the one-loop $E_{M1}$ [see Eq.~(\ref{defEM1})].
It contains additionally an iteration
of the one-loop potential $\delta^{\rm (1)} V$ and 
the term proportional to $\kappa^2$ from Eq.~(\ref{trafo2}):
\begin{widetext}
\begin{eqnarray}
\label{defEF1}
\lefteqn{E_{F1} = \langle \delta^{\rm (2)} V^i \rangle + 
2\, \left<\delta^{\rm (2)} V^i \,\frac{1}{(E-H)'}\,H_R\right>
+ \frac{F'^{\rm (2)}_1(0) + 2 F'^{\rm (2)}_2(0)}{8}\,\langle\vec{\nabla}^4V
+ 2\,{\rm i}\,\sigma^{ij}\,p^i\vec{\nabla}^2V\,p^j\rangle
+ F''^{\rm (2)}_1(0)\,\langle\vec{\nabla}^4V\rangle}
\nonumber \\ 
& & -\frac{F^{\rm (2)}_2(0)}{16}\,
\left<\Bigl\{\vec p^{\,2}, \vec{\nabla}^2V
+ 2\,\sigma^{ij}\,\nabla^i V\,p^j \Bigr\}\right>
+ \frac{F^{\rm (2)}_2(0)+\left[ F_2^{(1)}(0) \right]^2}{8}\,
\langle (\vec{\nabla} V)^2 \rangle
+\left<\,\delta^{\rm (1)} V\, \frac{1}{(E-H)'}\,
\delta^{\rm (1)} V\,\right>.
\end{eqnarray}
The two-loop form factors are given in Eq.~(\ref{ff})
below, and $\delta^{\rm (1)} V, \delta^{\rm (2)} V$ are
the one- and two-loop  components respectively of 
the potential given in Eq.~(\ref{defdeltaV}).
The explicit form of $\delta^{(2)}V^i$ can be found 
in Eq.~(\ref{d2Vi}) below.
                                                                  
$E_{F2}$ comes from the low-energy two-loop scattering amplitude
and is the analog of the one-loop  $E_{M2}$ in Eq.~(\ref{defEM2}).
The effective interaction is
\begin{equation}
\label{defH2L}
\delta^{\rm (2)} H = \chi^{(2)}\,(\vec{\nabla} V)^2 \,, 
\end{equation}
where $\chi^{(2)}$ is defined in Eq.~(\ref{eta2}) below.
It is assumed that vacuum polarization diagrams does not 
contribute in the current section  to form factors as well as to $\chi$.
The energy shift due to 
$\delta^{\rm (2)} H$ is 
\begin{equation}
\label{defEF2}
E_{F2} = \left< \delta^{\rm (2)} H\right> \,.
\end{equation}
It is a remarkable fact
that this two-loop scattering-amplitude contribution is infrared finite,
in contrast to the corresponding one-loop result in Eq.~(\ref{defEM2}).

For the two-loop problem, we redefine
$E_M$ to be the contribution where 
one of the photon momenta is of the order of
the electron mass, the second photon momentum is of order $(Z\,\alpha)^2$
and the electron momenta are of order $Z\,\alpha$. In the spirit of NRQED,
the contribution coming from large photon momenta is accounted for by
form factors. Therefore $E_M$ is given by the correction to Bethe logarithms
coming from one-loop form factors. It is a sum of two parts
\begin{equation}
\label{redefEM}
E_M = E_{M1}+E_{M2}\,.
\end{equation}
The contribution $E_{M1}$ 
is similar to the one-loop term $E_{L1}$
with $H_R$ replaced by $\delta^{\rm (1)} V$:
\begin{eqnarray}
\label{defEM12L}
E_{M1} &=& 
e^2\,\int \frac{d^d k}{(2\,\pi)^d\,2\,k}\,
\frac{d-1}{d}\,\delta_{\delta^{\rm (1)} V}\,
\left< \,\vec p\,\frac{1}{E-H-k}\,\vec p\, \right>\,.
\end{eqnarray}
We calculate it by splitting the integral in two parts $k<\Lambda$
and $k>\Lambda$ in analogy to the one-loop case,
\begin{eqnarray}
\label{EM12Lintermediate}
E_{M1} &=& e^2\, \int_0^\Lambda \frac{d^d k}{(2\,\pi)^d\,2\,k}\,
\frac{d-1}{d}\,\delta_{\delta^{\rm (1)} V}\,
\left< \, \vec p\,\frac{1}{E-H-k}\,\vec p\,\right>
\nonumber \\ 
&& +\frac{\alpha}{\pi}\, \frac{\xi}{2}\,
\left\{
\left< [\vec p,[\delta^{\rm (1)} V,\vec p]] \right> 
+ 2\,\left< \delta^{\rm (1)}V\,\frac{1}{(E-H)'}\,\nabla^2 V \right>
\right\}\,,
\end{eqnarray} 
where
\begin{equation}
\xi = \frac{1}{3\,\varepsilon} +
\left\{\frac{5}{9}-\frac{2}{3}\,\ln[2\,(Z\,\alpha)^2]\right\}
+\varepsilon\,\left\{
\frac{28}{27}-\frac23 \,\zeta(2)
+\frac{10}{9}\,\ln\left[\half\,(Z\,\alpha)^{-2}\right]
+\frac{2}{3}\,\ln^2\left[\half\,(Z\,\alpha)^{-2}\right]\right\}\,.
\end{equation}
We have not approached the limit $d=3$ in the first part, 
because $\delta^{\rm (1)} V$ contains $1/\varepsilon$. 
It will eventually cancel when combined with $E_L$,  and only then
one approaches this limit.
$E_{M2}$ is similar to the one-loop $E_{L3}$
and comes from the $F_2(0)$-correction to the coupling with the radiation field,
\begin{equation}
H_{FW} =-\frac{e}{4}\,\sigma^{ij}\,B^{ij}\,[1+F_2(0)] - \frac{e}{8}\,
[1+2\,F_2(0)][\nabla\cdot\vec E+\sigma^{ij}\,(E^i\,\pi^j+\pi^j\,E^i)]\,,
\end{equation}   
which yields
\begin{equation}
\delta j^i = \frac{F_2^{(1)}(0)}{2}\,\sigma^{ik}\,(\vec k\cdot\vec r\,k^k
+ {\rm i}\,k\,p^k-\nabla^k V) \simeq
- \frac{\alpha}{\pi}\,\frac{2\,d-3}{4\,(d-1)}\,\sigma^{ik}\,\nabla^k V\,.
\end{equation}
The corresponding correction $E_{M2}$ is
\begin{eqnarray}
E_{M2} &=& 2\,e^2\,\int \frac{d^d k}{(2\,\pi)^d\,2\,k}\,
\left(\delta^{ij}-\frac{k^i\,k^j}{k^2}\right)\,
\left< \delta j^i\frac{1}{E-H-k}\,p^j \right> \nonumber \\
&=& -\frac{1}{2}\,\left(\frac{\alpha}{\pi}\right)^2\,\int_0^\Lambda dk\,k\,
\left<\,\sigma^{ij}\,\nabla^j V
\frac{1}{E-H-k}\,p^i\right> \,,
\end{eqnarray}
and this integral in $(Z\,\alpha)^6$ order does not depend on the cut-off 
in the limit $\lambda\rightarrow\infty$, when one drops the linear 
term in $\lambda$.

The low-energy part $E_L$, appropriately redefined 
for the two-loop problem, is a contribution from two low-energy 
photon momenta, $k_i\sim (Z\,\alpha)^2$. 
Its explicit expression is rather long:
\begin{align}
\label{redefEL}
& E_L =
\left[e^2\,\int 
\frac{d^d k_1}{(2\,\pi)^d\,2\,k_1}\,\frac{d-1}{d}\right]\,
\left[e^2\,
\int \frac{d^d k_2}{(2\,\pi)^d\,2\,k_2}\,\frac{d-1}{d}\right]\,
P(k_1, k_2) \,, \nonumber \\ 
& P(k_1,k_2) = 
\left< p^i\,\frac{1}{E-(H+k_1)}\,p^j\,
\frac{1}{E-(H+k_1+k_2)}\,p^i\,\frac{1}{E-(H+k_2)}\,
p^j \right> \nonumber \\
& \quad + \frac{1}{2}\,\left< p^i\,\frac{1}{E-(H+k_1)}\,p^j\,
\frac{1}{E-(H+k_1+k_2)}\,p^j\,\frac{1}{E-(H+k_1)}\,
p^i \right> \nonumber \\ 
& \quad + \frac{1}{2}\,\left< p^i\,\frac{1}{E-(H+k_2)}\,p^j\,
\frac{1}{E-(H+k_1+k_2)}\,p^j\,\frac{1}{E-(H+k_2)}\,
p^i \right> \nonumber \\
& \quad + \left< p^i\,\frac{1}{E-(H+k_1)}\,p^i\,
\frac{1}{(E-H)'}\,p^j\,\frac{1}{E-(H+k_2)}\,
p^j \right> \nonumber \\
& \quad - \frac{1}{2}\,\left< p^i\,\frac{1}{E-(H+k_1)}\,p^i \right>\,
\left< p^j\,\frac{1}{[E-(H+k_2)]^2}\,p^j \right>
- \frac{1}{2}\,\left< p^i\,\frac{1}{E-(H+k_2)}\,p^i \right>\,
\left< p^j\,\frac{1}{[E-(H+k_1)]^2}\,p^j \right>
\nonumber \\ 
& \quad + \left< p^i\frac{1}{E-(H+k_1)} \frac{1}{E-(H+k_2)}p^i\right> 
- \frac{1}{k_1+k_2}\left< p^i\frac{1}{E-(H+k_2)} p^i\right> 
- \frac{1}{k_1+k_2}\left< p^i\frac{1}{E-(H+k_1)} p^i \right>.
\end{align}
We calculate $E_L$ by splitting both integrals in a way
similar to the derivation presented in~\cite{Pa2001}, 
\begin{eqnarray}
E_L &=& 
\left(\frac{\alpha}{\pi}\right)^2\,\frac{(Z\,\alpha)^6}{n^3}\,b_L 
+e^2\,\int^{\Lambda}_0\, \frac{d^d k_2}{(2\,\pi)^d\,2\,k_2}\,
\frac{d-1}{d}\,
\frac{\alpha}{\pi}\,\frac{\xi}{2}\,\delta_{\nabla^2 V}\,
\left< \,\vec p\,\frac{1}{E-H-k}\,\vec p\, \right>
\nonumber \\ &&
+ \left[\frac{\alpha}{\pi}\,\frac{\xi}{2}\right]^2\,
\left[\left< \vec{\nabla}^2 V \,
\frac{1}{(E-H)'}\,
\vec{\nabla}^2 V\right> 
+ \frac{1}{2} \, \left< \vec{\nabla}^4 V\right> \right]\,.
\end{eqnarray}
Here, the two-loop Bethe logarithm  $b_L$ is obtained 
as the finite part of the integral
\begin{equation}
\frac{(Z\,\alpha)^6}{n^3}\,b_L = 
\frac{4}{9}\, \int_0^{\Lambda_1} dk_1\,k_1\,
\int_0^{\Lambda_2} dk_2\,k_2\,P(k_1, k_2)\,,
\end{equation}
where it is assumed that the following limits
are performed in order: first $d\rightarrow 3$, next
$\lambda_2\rightarrow\infty$ and finally 
$\lambda_1\rightarrow\infty$ in the above.
This definition of $b_L$ corresponds to the one 
in Refs.~\cite{PaJe2003,Je2004b60}. 

%
% Two-loop result
%
\subsection{General result for the pure two-loop self-energy}
\label{calciRES}

The pure two-loop self-energy contribution up to the order
$\alpha^2\,(Z\,\alpha)^6$, denoted $\delta^{(2)} E^i$
(see Fig.~\ref{fig1}),
may now be obtained as the sum of $E_F + E_M + E_L$.
With the partial results
given in  Eqs.~(\ref{defEF}), (\ref{redefEM}) and (\ref{redefEL}),
respectively, we obtain
\begin{align}
\label{deltaEi}
& \delta^{(2)} E^i = \left< \delta^{(2)} V^i \right> +
\left(\frac{\alpha}{\pi}\right)^2\,
\frac{(Z\,\alpha)^6}{n^3}\,\left[b_L+
\left(\frac{10}{9}
+ \frac{4}{3} \ln\left[\half\,(Z\,\alpha)^{-2}\right] \right)\,N
+ \beta_{4} + \beta_{5}\right]
+ \left< V_I\frac{1}{(E-H)'}\,V_I \right>
\nonumber \\ 
& + 2\,\left< \delta^{(2)} V^i \,\frac{1}{(E-H)'}\,H_R \right> 
+\left(\frac{\alpha}{\pi}\right)^2\,
\left[\frac{31}{256} 
+ \frac{3}{16} \,\zeta(2)\,\ln (2)
- \frac{5}{32} \,\zeta(2)
- \frac{3}{64}\,\zeta(3)\right]\,
\left< 
\Bigl\{p^2, \vec{\nabla}^2 V + 2\,\sigma^{ij}\,\nabla^i V\,p^j\Bigr\}
\right> 
\nonumber \\ 
& +\left(\frac{\alpha}{\pi}\right)^2\,
\left[-\frac{559}{1152}  
+ \frac{17}{8} \zeta(2)\,{\ln(2)}
- \frac{41}{72}\,\zeta(2) 
- \frac{17}{32}\,\zeta(3) \right]\,
\langle(\vec{\nabla} V)^2\rangle
\nonumber \\ 
& +\left(\frac{\alpha}{\pi}\right)^2\,
\left[-\frac{3295}{41472} 
+ \frac{9}{10}\zeta(2)\,\ln (2)
- \frac{4063}{14400}\zeta(2)
- \frac{9}{40}\zeta(3)
+ \frac{5}{54}\ln\left[\half\,(Z\,\alpha)^{-2}\right]
+ \frac{1}{18}\ln^2\left[\half\,(Z\,\alpha)^{-2}\right]
\right]\,
\langle\vec{\nabla}^4 V\rangle 
\nonumber \\ 
& + \left(\frac{\alpha}{\pi}\right)^2\,
\left[-\frac{3059}{23040}  
- \frac{1}{5}\, \zeta(2)\,\ln(2)
+ \frac{1321}{5760} \, \zeta(2)
+ \frac{1}{20}\,\zeta(3)
+ \frac{1}{24}\, \ln\left[\half\,(Z\,\alpha)^{-2}\right]
\right]\,
\langle 2\,{\rm i}\,\sigma^{ij}\, p^i\,\vec{\nabla}^2V\,p^j\rangle.
\end{align}
Here, the first term $\langle \delta^{(2)} V^i\rangle$ is of
lower-order [$\alpha^2 (Z\alpha)^4$], and
\begin{eqnarray}
V_I &=&\frac{\alpha}{\pi}\,\left[
\frac{\vec{\nabla}^2 V}{4} \, 
\left(\frac{10}{9} \,
+\frac{4}{3}\ln\left[\half\,(Z\,\alpha)^{-2}\right]\right)
+\frac{\sigma^{ij}}{4}\,\nabla^i\,V\,p^j\right] \,.\\
\label{d2Vi}
\delta^{(2)} V^i &=& 
\left(\frac{\alpha}{\pi}\right)^2\,
\left\{\left[ -\frac{163}{288}
+ \frac{9}{4}\,\zeta(2)\,\ln 2
- \frac{85}{144}\,\zeta(2)
- \frac{9}{16}\,\zeta(3)\right]\, \vec{\nabla}^2 V 
\right. \nonumber \\[0.5ex]
&& \left. + \left[-\frac{31}{32}
- \frac{3}{2}\,\zeta(2)\,\ln 2
+ \frac{5}{4}\,\zeta(2)
+ \frac{3}{8}\,\zeta(3)
\right]\,
\sigma^{ij}\,\nabla^i V\,p^j\right\}\,.
\end{eqnarray}
The various generalized Bethe logarithms that enter into 
Eq.~(\ref{deltaEi}), are given as follows
(with the implicit assumption that polynomial divergences
as well as logarithmic ones for large $\lambda = \Lambda/(Z\,\alpha)^2$ 
are dropped)
\begin{subequations}
\begin{eqnarray}
\label{defN}
\frac{(Z\,\alpha)^6}{n^3}\,N &=& 
\frac{2}{3}\, Z\,\alpha\, \int_0^{\Lambda}\, dk\,k\,
\delta_{\pi\,\delta^3(r)}
\left< \,\vec p\,\frac{1}{E-H-k}\,\vec p\,\right> \,, \\[0.5ex]
\label{defbeta4}
\frac{(Z\,\alpha)^6}{n^3}\,\beta_{4} &=& 
\frac{2}{3}\, \int_0^{\Lambda}dk\,k\,
\delta_{(\sigma^{ij}\nabla^i\,V\,p^j/4)}
\left< \,\vec p\,\frac{1}{E-H-k}\,\vec p\,\right> \,, \\[0.5ex]
\label{defbeta5}
\frac{(Z\,\alpha)^6}{n^3}\,\beta_{5} &=& 
\frac{2}{3}\, \int_0^\Lambda dk\,k\,
\left< -\frac34\,\sigma^{ij}\nabla^j V\frac{1}{E-H-k}\,p^i \right>\,,
\end{eqnarray}
\end{subequations}
The $N$ term has previously been defined in
Refs.~\cite{Pa2001,PaJe2003}; it is generated by a Dirac delta correction 
to the Bethe logarithm.
All the explicit matrix element occurring in
the formula (\ref{deltaEi}) can be calculated using standard techniques,
for arbitrary hydrogenic states with nonvanishing angular momentum,
and for the normalized difference (\ref{assumptions}) of $S$ states.
The evaluation of the generalized Bethe logarithms $N$, 
$\beta_4$, and $\beta_5$ is more complicated 
(see Refs.~\cite{Je2003jpa,JeEtAl2003}).
The calculation of the two-loop Bethe logarithm $b_L$ for 
arbitrary excited hydrogenic states is a challenging numerical problem.
So far results have been obtained only for excited $S$
states~\cite{PaJe2003,Je2004b60}.
The formula (\ref{deltaEi}) thus provides the basis for 
complete two-loop calculations in the order $\alpha^2\,(Z\,\alpha)^6$,
and reduces the remaining part of the problem, for a general 
hydrogenic state, to a well-defined and in 
essence merely technical numerical calculations.
In the following sections we discuss the evaluation of the formula
(\ref{deltaEi}) for particular hydrogenic
states for which the generalized Bethe logarithms can be inferred
from previous calculations. These comprise the
fine-structure difference of $D$ and $P$ states, 
and the normalized difference for $S$ states.

\begin{table}[htb]
%\begin{minipage}{18.0cm}
\caption{\label{tabledi} Numerical values for the pure two-loop self-energy 
diagrams as shown in Fig.~\ref{fig1}. The $B_{60}$-coefficients receive 
a superscript $i$.}
\begin{tabular}{r@{\hspace{1.0cm}}.@{\hspace{1.0cm}}.....}
\hline
\hline
\rule[-3mm]{0mm}{8mm}
$n$ &
\multicolumn{1}{c}{$B^i_{60}(D_{5/2}-D_{3/2})$} & 
\multicolumn{1}{c}{$B^i_{60}(P_{3/2}-P_{1/2})$} &
\multicolumn{1}{c}{$b_L(nS)$} & 
\multicolumn{1}{c}{$N(nS)$} & 
\multicolumn{1}{c}{$R(n)$} &
\multicolumn{1}{c}{$B^i_{60}(nS) - B^i_{60}(1S)$} \\
\hline
1 & \empt       & \empt       & -81x.4(3) & 17x.855\,672\,03(1) &   \empt       &   \empt     \\
2 & \empt       & -0x.361~196 & -66x.6(3) & 12x.032\,141\,58(1) &  -0x.671\,347 &   14x.1(4)  \\
3 & -0x.018~955 & -0x.410~149 & -63x.5(6) & 10x.449\,809(1)     &  -1x.041\,532 &   16x.9(7)  \\
4 & -0x.022~253 & -0x.419~927 & -61x.8(8) & 9x.722\,413(1)      &  -1x.254\,980 &   18x.3(10) \\
5 & -0x.023~395 & -0x.420~828 & -60x.6(8) & 9x.304\,114(1)      &  -1x.392\,573 &   19x.4(11) \\
6 & -0x.023~826 & -0x.419~339 & -59x.8(8) & 9x.031\,832(1)      &  -1x.488\,456 &   20x.1(11) \\
\hline
\hline
\end{tabular}
%\end{minipage}
\end{table}

%
% Fine-structure difference of $D$ states
%
\subsection{Results for the fine-structure difference of $D$ states}
\label{twlpDfs}

For $D$ states, we use the general result (\ref{deltaEi}) and the 
fact that matrix elements involving a Dirac-delta function vanish.
Thus, logarithmic terms for $D$ levels vanish,
$B^i_{61}(D_{3/2}) = B^i_{61}(D_{5/2}) = 0$.
The absence of logarithmic terms even holds for the sum of all
two-loop diagrams (not only for the subset $i$), and even for   
arbitrary states with orbital angular momentum $l > 2$.
This result generalizes the well-known fact 
that the double-logarithmic contribution 
$B_{62}$ vanishes for states with $l\geq 2$~\cite{Ka1996,JeNa2002}. 
For the fine-structure difference of $B^i_{60}$,
we use the result in Eq.~(\ref{deltaEi}) and the matrix 
elements in Eq.~(\ref{matelemD}), to obtain
\begin{align}
\label{fsDi}
& B^i_{60}(D_{5/2}-D_{3/2}) =
-\frac{38497}{403200} 
- \frac{133}{640\,n} 
+ \frac{895}{1344\,n^2} 
+ \left( -\frac{3817}{25200}
- \frac{13}{40\,n}
+ \frac{29}{28\,n^2} \right) \, \zeta(2)\,\ln(2)
\nonumber\\[2ex]
& + \left( \frac{3817}{30240} 
+ \frac{13}{48\,n}
- \frac{145}{168\,n^2} \right) \, \zeta(2) 
+ \left( \frac{3817}{100800} 
+ \frac{13}{160\,n} 
- \frac{29}{112\,n^2} \right) \,\zeta(3) 
+ \beta_{4}(D_{5/2}-D_{3/2}) 
+ \beta_{5}(D_{5/2}-D_{3/2})\,.
\end{align}
Numerical data for $B_{60}(D_{5/2}-D_{3/2})$
can be found in Tab.~\ref{tabledi}. The unknown 
two-loop Bethe logarithm $b_L(nD)$ 
does not contribute to the fine-structure difference of
$D$ states.

%
% Fine-structure difference of $P$ states
%
\subsection{Results for the fine-structure difference of $P$ states}
\label{twlpPfs}

We again use the fact that the 
unknown two-loop Bethe logarithm $b_L$ does not 
contribute to the fine-structure difference of $P$ 
states. With the help of the 
general result in Eq.~(\ref{deltaEi}) and the matrix
elements in Eq.~(\ref{matelemP}), we obtain
\begin{equation}
B_{61}(P_{3/2}-P_{1/2}) = -\frac{1}{3}\,\left( 1 - \frac{1}{n^2}\right)
\end{equation}
in agreement with the literature~\cite{JePa2002} and
\begin{align}
\label{fsPi}
& B^i_{60}(P_{3/2}-P_{1/2}) =
- \frac{217}{1280} 
- \frac{151}{128\,n}
+ \frac{325}{288\,n^2} 
+ \left( \frac{1}{3} - \frac{1}{3\,n^2} \right) \,\ln(2) 
+ \left( -\frac{103}{240} 
- \frac{15}{8\,n} 
+ \frac{37}{20\,n^2} \right) \,
\zeta(2)\,\ln(2)
\nonumber\\[2ex]
& + \left( -\frac{23}{160} 
+ \frac{25}{16\,n} 
- \frac{749}{720\,n^2} \right) \,\zeta(2) 
+ \left( \frac{103}{960} 
+ \frac{15}{32\,n} 
- \frac{37}{80\,n^2} \right) \,\zeta(3) 
+ \beta_{4}(P_{3/2}-P_{1/2}) 
+ \beta_{5}(P_{3/2}-P_{1/2}) \,.
\end{align}
Numerical values of the relevant quantities
for $n = 2,\dots,6$ can be found in Tab.~\ref{tabledi}.
They are in full agreement with results previously 
obtained in~\cite{JePa2002}.
The generalized Bethe logarithms $\beta_{4}$ and 
$\beta_{5}$ in these expressions are equivalent to the 
quantities $\Delta_{\mathrm{fs}}{{\ell}}_4(n)$ and 
$\Delta_{\mathrm{fs}}{{\ell}}_5(n)$ as defined in Ref.~\cite{JePa2002}.
In the context of the current 
investigation, the numerical values of $\Delta_{\mathrm{fs}}{{\ell}}_4(n)$ and
$\Delta_{\mathrm{fs}}{{\ell}}_5(n)$ were reevaluated with improved accuracy 
as compared to~Ref.~\cite{JeKePa2002} and the data in 
Tab.~\ref{tabledi} are consistent with them.

%
% Normalized difference of $S$ states
%
\subsection{Results for the normalized difference of $S$ states}
\label{twlpSnormalized}

We evaluate the general formula
given in Eq.~(\ref{deltaEi}) for the normalized difference 
of $S$ states, using the matrix elements given in Eq.~(\ref{matelemS}).
In the result, we identify terms with the square of the logarithm 
$\ln[(Z\,\alpha)^{-2}]$ ($B^i_{62}$ coefficient), and with single 
logarithm ($B^i_{61}$ coefficient), and the 
nonlogarithmic term $B^i_{60}$. The results discussed here
probably are the phenomenologically most important ones
reported in this paper,
because of the high accuracy of two-photon spectroscopic
experiments which involve $S$--$S$ transitions.

For the double-logarithmic term, we recover the following known result
(see~Refs.~\cite{Pa2001,Je2003plb}),
\begin{eqnarray}
\label{B62}
B^i_{62}(nS) - B^i_{62}(1S)
&=& \frac{16}{9} \left( \frac34 + \frac{1}{4 n^2} - 
\frac1n + \gamma - \ln(n) \! + \! \Psi(n) \right)\,.
\end{eqnarray}
Here $\Psi$ denotes the logarithmic derivative of the gamma function,
and $\gamma = 0.577216\dots$ is Euler's constant.
The result for $B_{61}$, restricted to the
two-loop diagrams in Fig.~\ref{fig1},
reads~\cite{Pa2001,Je2003plb}
\begin{equation}
\label{ndepB61i}
B^i_{61}(nS) - B^i_{61}(1S)=
\frac43 \, \left[ N(nS) - N(1S) \right]
+ \left( \frac{80}{27} - \frac{32}{9} \, \ln 2\right)\,
\left(\frac34 - \frac1n + \frac{1}{4 n^2} 
+ \gamma - \ln(n) + \Psi(n) \right)\,.
\end{equation}
This result is recovered
here from Eq.~(\ref{deltaEi}), using the matrix elements in
Eq.~(\ref{matelemS}). Moreover, we obtain the 
complete $n$-dependence of the nonlogarithmic term $B^i_{60}$:
\begin{align}
\label{ndepB60i}
& B^i_{60}(nS) - B^i_{60}(1S) =
b_L(nS) - b_L(1S) 
+ \left( \frac{10}{9} - \frac43\,\ln(2) \right) \, [N(nS) - N(1S)]  
\nonumber\\
& + \frac{10529}{5184} - \frac{14099}{2592\,n} + \frac{17699}{5184\,n^2} 
+ \left( \frac{4}{3} 
- \frac{16}{9\,n}
+ \frac{4}{9\,n^2} \right) \, \ln^2 (2)
+ \left( -\frac{20}{9} 
+ \frac{80}{27\,n}
- \frac{20}{27\,n^2} \right) \,\ln(2) 
\nonumber\\
& + \left( - \frac{53}{15}
+ \frac{35}{2\,n}
- \frac{149}{30\,n^2} \right) \, \zeta(2)\,\ln(2) 
+ \left( \frac{1357}{2700} 
- \frac{167}{36\,n} 
+ \frac{2792}{675\,n^2} \right) \, \zeta(2) 
+ \left( \frac{53}{60} 
- \frac{35}{8\,n}
+ \frac{419}{120\,n^2} \right) \,\zeta(3)  
\nonumber\\
& + \left( - \frac{497}{1296} 
+ \frac{16}{9}\ln^2(2)
- \frac{80}{27}\ln (2)
+ 8\,\zeta(2)\ln(2) 
- \frac{79}{36}\zeta(2)
- 2\,\zeta(3)\right) \,
\left[ \gamma + \Psi(n) - \ln(n) \right]\,.
\end{align}
The generalized Bethe logarithms
$\beta_4$ and $\beta_5$, which make an occurrence in Eq.~(\ref{deltaEi}) 
but are not present in Eq.~(\ref{ndepB60i}), vanish for $S$ states. 
The result (\ref{ndepB60i}) can also be written as
\begin{equation}
\label{defRnS}
B^i_{60}(nS) - B^i_{60}(1S) = b_L(nS) - b_L(1S) + R(n)\,,
\end{equation}
which provides a definition of the remainder $R(n)$.
Numerical values for $b_L(nS)$, $N(nS)$, $R(n)$ and the 
normalized $S$-state difference $B^i_{60}(nS) - B^i_{60}(1S)$ are given in 
Table~\ref{tabledi}, and we have the opportunity to correct
a calculational error for $N(2S)$ whose 
value had previously been given as $12.032209$
in~\cite{Je2003jpa}. 

\begin{figure}[htb]
\begin{minipage}{8.6cm}
\centerline{\includegraphics[width=0.5\linewidth]{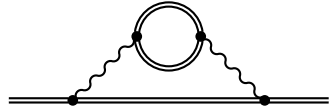}}
\caption{\label{fig2}
Feynman diagram with a vacuum-polarization loop
in the self-energy virtual photon line
(this single diagram forms subset $ii$ in the convention
adopted in this paper).}
\end{minipage}
\end{figure}

%
% FERMION LOOP IN THE SELF-ENERGY PHOTON LINE
%
\section{FERMION LOOP IN THE SELF-ENERGY PHOTON LINE}
\label{calcii}

We here calculate the mixed self-energy vacuum-polarization diagram in 
Fig.~\ref{fig2}. The result can be 
easily inferred from the terms in square brackets 
in Eqs.~(\ref{defH2L}),~(\ref{ff}), 
and~(\ref{eta1plus2}), and reads
\begin{align}
\label{deltaEii}
& \delta^{(2)} E^{ii} = \left< \delta^{(2)} V^{ii} \right> +
2\,\left< \delta^{(2)} V^{ii}\,\frac{1}{(E-H)'}\,H_R \right> 
+\left(\frac{\alpha}{\pi}\right)^2\,
\left[ -\frac{119}{576} + \frac{1}{8} \, \zeta(2) \right]\,
\left< 
\Bigl\{p^2, \vec{\nabla}^2 V + 2\,\sigma^{ij}\,\nabla^i V\,p^j\Bigr\}
\right> 
\nonumber \\ 
& +\left(\frac{\alpha}{\pi}\right)^2\,
\left\{
\left[ \frac{119}{288} - \frac{1}{4} \, \zeta(2) \right]
\langle(\vec{\nabla} V)^2\rangle
+
\left[ -\frac{4511}{51840} + \frac{65}{1152} \, \zeta(2)
\right]
\langle\vec{\nabla}^4 V\rangle 
+ 
\left[ \frac{2633}{10368} - \frac{175}{1152} \, \zeta(2) \right]\,
\langle 2\,{\rm i}\,\sigma^{ij}\, p^i\,\vec{\nabla}^2V\,p^j\rangle
\right\},
\end{align}
where 
\begin{align}
\label{defdeltaVii}
\delta^{(2)} V^{ii} = 
\left(\frac{\alpha}{\pi}\right)^2\,
\left\{\left[ -\frac{7}{324}  + \frac{5}{144} \, \zeta(2)
\right]\, \vec{\nabla}^2 V 
+ \left[\frac{119}{72} - \zeta(2) \right]\,
\sigma^{ij}\,\nabla^i V\,p^j\right\}\,.
\end{align}
is a radiative potential in the sense of Eq.~(\ref{defdeltaV}), but 
includes here only the vacuum polarization part of form factors.
We observe the absence of $\ln (Z\,\alpha)$ terms.

\begin{table}[htb]
\begin{minipage}{12.0cm}
\caption{\label{tabledii} Values of
the fine-structure for $D$ and $P$ states and normalized difference of $S$
states coming from from fermion loop  diagrams in Fig.~\ref{fig2}.}
\begin{tabular}{r@{\hspace{1.0cm}}.@{\hspace{1.0cm}}.@{\hspace{1.0cm}}.}
\hline
\hline
\rule[-3mm]{0mm}{8mm}
$n$ &
\multicolumn{1}{c}{$B^{ii}_{60}(D_{5/2} - D_{3/2})$} &
\multicolumn{1}{c}{$B^{ii}_{60}(P_{3/2} - P_{1/2})$} & 
\multicolumn{1}{c}{$B^{ii}_{60}(nS)-B^{ii}_{60}(1S)$} \\
\hline
2 &        \empt       &       -0x.013\,435 & 0x.109\,999 \\
3 &        0x.000\,757 &       -0x.017\,089 & 0x.114\,502 \\
4 &        0x.000\,878 &       -0x.018\,613 & 0x.110\,743 \\
5 &        0x.000\,915 &       -0x.019\,431 & 0x.106\,566 \\
6 &        0x.000\,925 &       -0x.019\,935 & 0x.102\,982 \\
\hline
\hline
\end{tabular}
\end{minipage}
\end{table}

Numerical values for $S$, $P$, and $D$ states can now be obtained
using matrix elements in Eqs.~(\ref{matelemD}), (\ref{matelemP}) 
and~(\ref{matelemS}). For the fine-structure intervals, 
we obtain
\begin{align}
\label{fsDii}
B^{ii}_{60}(D_{5/2} - D_{3/2}) =&
\frac{64889}{388800} 
+ \frac{1547}{4320\,n}  
- \frac{493}{432\,n^2} 
+ \left( -\frac{3817}{37800} 
- \frac{13}{60\,n} 
+ \frac{29}{42\,n^2} 
\right) \, 
\zeta(2)\,,
\\
\label{fsPii}
B^{ii}_{60}(P_{3/2} - P_{1/2}) =&
\frac{5293}{25920} 
+ \frac{595}{288\,n} 
- \frac{2867}{1620\,n^2} 
+ \left( - \frac{11}{80} 
- \frac{5}{4\,n} 
+ \frac{781}{720\,n^2} \right) \,\zeta(2)\,.
\end{align}
Considering $S$ states, as is evident from Eq.~(\ref{deltaEii}), 
using the matrix elements in
Eq.~(\ref{matelemS}), the normalized difference of $B^{ii}_{61}$ 
vanishes, $B^{ii}_{61}(nS) - B^{ii}_{61}(1S) = 0$,
and this result is in agreement with the literature.
For the normalized $n$-dependence of $B^{ii}_{60}$, we obtain the following 
result,
\begin{align}
\label{ndepB60ii}
B^{ii}_{60}(nS) - B^{ii}_{60}(1S) =&
- \frac{21319}{6480} + \frac{1015}{648\,n} + \frac{1241}{720\,n^2}  
+ \left( \frac{301}{144} - \frac{31}{36\,n} 
- \frac{59}{48\,n^2} \right) \, \zeta(2) \nonumber \\
& + \left[ \frac{1099}{324} - \frac{77}{36}\,\zeta(2) \right] \,
\left[ \gamma + \Psi(n)  - \ln (n) \right]\,.
\end{align}
Numerical values are presented in  
Tab.~\ref{tabledii}.

%
% COMBINED SELF-ENERGY WITH A FERMION LOOP IN THE COULOMB PHOTON LINE
%
\section{COMBINED SELF-ENERGY WITH A FERMION LOOP IN THE COULOMB PHOTON LINE}
\label{calciii}

The Feynman diagrams in Fig.~\ref{fig3} 
represent the modification of a leading one-loop self-energy correction
by  a perturbing Uehling potential
$V_U = -\frac{4}{15}\,\alpha \, (Z\alpha)\, \delta^d(r)
= -\frac{\alpha}{15 \, \pi} \,\vec{\nabla}^2 V$.
One can easily obtain the result from $E_{L0}+E_{M0}$ in 
Eqs.~(\ref{EL0}) and (\ref{EM0}),
by replacing the Coulomb potential  $V$ by $V+ V_U$, and
expanding all matrix elements in $V_U$, up to the linear 
terms. The result is
\begin{align}
\label{deltaEiii}
\delta^{(2)} E^{iii} =& 
-\frac{4}{15} \, \left(\frac{\alpha}{\pi}\right)^2 \,
\frac{(Z\,\alpha)^6}{n^3} \, N
- \frac{1}{15} \, \left(\frac{\alpha}{\pi}\right)^2 \,
\left[ \frac59 + \frac23 \, \ln\left[\half \,(Z\alpha)^{-2}\right] \right]\,
\left< \vec{\nabla}^2 V \, \left( \frac{1}{E-H} \right)' \,
\vec{\nabla}^2 V \right>
\nonumber\\
&- \frac{1}{120} \, \left(\frac{\alpha}{\pi}\right)^2 \,
\langle 2\,{\rm i}\,\sigma^{ij}\, p^i\,\vec{\nabla}^2V\,p^j\rangle
%
% - \frac{1}{15} \, \left(\frac{\alpha}{\pi}\right)^2 \,
% \left< \vec{\nabla}^4 V \right> \,
% \left[ \frac{5}{18} + \frac13 \, 
% \ln\left[\half \,(Z\alpha)^{-2}\right] \right]\,.
- \left(\frac{\alpha}{\pi}\right)^2 \,
\left[ \frac{1}{54} + \frac{1}{45} \, 
\ln\left[\half \,(Z\alpha)^{-2}\right] \right]\,
\left< \vec{\nabla}^4 V \right> \,.
\end{align}
\end{widetext}
Here, $N$ is a correction to the Bethe logarithm 
as defined in Eq.~(\ref{defN}).
All matrix elements in this result vanish 
for $D$-states and for states with higher angular momenta.
The absence of both logarithmic as well as nonlogarithmic terms holds 
from subset $iii$ holds for 
arbitrary states with orbital angular momentum $l > 2$.
For $P$ states, we obtain the fine-structure 
difference
$B^{iii}_{61}(nP_{3/2}) - B^{iii}_{61}(nP_{1/2}) = 0$,
in agreement with the literature. For the nonlogarithmic term,
we obtain
\begin{equation}
\label{fsPiii}
B^{iii}_{60}(nP_{3/2}) - B^{iii}_{60}(nP_{1/2}) 
= \frac{1}{15} \, \left(1 - \frac{1}{n^2} \right)\,.
\end{equation}
As a last example, we consider the $S$-state normalized difference 
defined in Eq.~(\ref{assumptions}).
using matrix elements given in (\ref{matelemS}).
For the double-logarithmic term, we recover the known 
result $B^{iii}_{62}(nS) = 0$
(see~Refs.~\cite{Pa2001,Je2003plb}).
The result for $B^{iii}_{61}$
reads~\cite{Pa2001,Je2003plb}
\begin{eqnarray}
\label{ndepB61iii}
\lefteqn{B^{iii}_{61}(nS) - B^{iii}_{61}(1S) }
\nonumber\\[2ex]
&=& - \frac{32}{45}  
\left(\frac34 + \frac{1}{4 n^2} - \frac1n + \gamma + \Psi(n) - \ln(n) 
\right) \,.
\end{eqnarray}
This result is recovered here from Eq.~(\ref{deltaEiii}).
As a new result, we obtain the 
complete $n$-dependence of the nonlogarithmic term 
$B^{iii}_{60}$:
\begin{align}
\label{ndepB60iii}
& B^{iii}_{60}(nS) - B^{iii}_{60}(1S)
= -\frac{4}{15}\, \left[ N(nS) - N(1S) \right]
\nonumber\\[2ex]
& - \frac{4}{9} + \frac{16}{27 n} - \frac{4}{27 n^2} 
+ \left( \frac{8}{15} 
- \frac{32}{45 n}
+ \frac{8}{45 n^2} \right) \ln(2) 
\nonumber\\
& + \left( -\frac{16}{27} 
+ \frac{32}{45}\,\ln (2) \right) \,
\left[ \gamma + \Psi(n) - \ln(n) \right]\,.
\end{align}
Using this formula, it is then possible to 
infer the values of $B^{iii}_{60}(nS) - 
B^{iii}_{60}(1S)$ as given in Tab.~\ref{tablediii}.

\begin{figure}[htb]
\begin{minipage}{8.6cm}
\centerline{\includegraphics[width=0.7\linewidth]{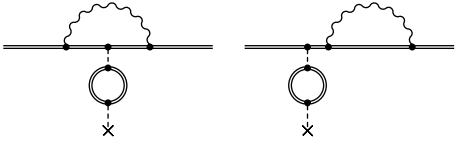}}
\caption{\label{fig3}
Two-loop diagrams (subset $iii$) generated by a 
fermion loop in the Coulomb exchange of a one-loop self-energy.}
\end{minipage}
\end{figure}

\begin{table}[htb]
\begin{minipage}{8.6cm}
\caption{\label{tablediii} Values of the difference
$B^{iii}_{60}(nS) - B^{iii}_{60}(1S)$ for the diagrams
in subset $iii$.}
\begin{tabular}{r@{\hspace*{0.6cm}}.}
\hline
\hline
\rule[-3mm]{0mm}{8mm}
$n$ & 
\multicolumn{1}{c}{$B^{iii}_{60}(nS) - B^{iii}_{60}(1S)$} \\
\hline
2 &      1.491\,199 \\
3 &      1.890\,577 \\
4 &      2.072\,903 \\
5 &      2.177\,348 \\
6 &      2.245\,177 \\
\hline
\hline
\end{tabular}
\end{minipage}
\end{table}

%
% PURE TWO-LOOP VACUUM POLARIZATION: subset iv
%
\section{PURE TWO-LOOP VACUUM POLARIZATION}
\label{calciv}

\begin{figure}[htb]
\begin{minipage}{8.6cm}
\centerline{\includegraphics[width=0.7\linewidth]{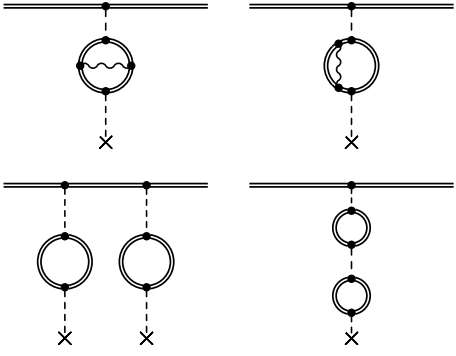}}
\caption{\label{fig4}
The remaining two-loop diagrams (subset $iv$) involve at least
one closed fermion loop in the Coulomb photon exchange between 
electron and nucleus, and no self-energy photons.}
\end{minipage}
\end{figure}

We investigate the subset of Feynman diagrams in 
Fig.~\ref{fig4}.
The vacuum polarization correction to the Coulomb potential is
\begin{align}
& -\frac{4\pi \,Z\,\alpha}{\vec{q}^{\,2}} \to 
-\frac{4\pi \,Z\,\alpha}{\vec{q}^{\,2}}\,
\frac{1}{\left( 1 + {\bar \omega}(-\vec{q}^{\,2}) \right)} 
\nonumber\\
& =-\frac{4\pi \,Z\,\alpha}{\vec{q}^{\,2}}\,
\left[1- {\bar \omega}(-\vec{q}^{\,2}) + 
{\bar \omega}(-\vec{q}^{\,2})^2 + \ldots\right] \,,
\end{align}
where the one- and two-loop parts read~\cite{Sc1970,BaRe1973}
as follows,
\begin{align}
{\bar \omega}(-\vec{q}^{\,2}) =& {\bar \omega}^{(1)}(-\vec{q}^{\,2}) + 
{\bar \omega}^{(2)}(-\vec{q}^{\,2})+\,\ldots, \nonumber\\
{\bar \omega}^{(1)}(-\vec{q}^{\,2}) =& \left(\frac{\alpha}{\pi}\right)\,
(-\vec{q}^{\,2}) \, \left( \frac{1}{15} 
- \frac{\vec{q}^{\,2}}{140} +\ldots \right)\,,\\
{\bar \omega}^{(2)}(-\vec{q}^{\,2}) =& \left(\frac{\alpha}{\pi}\right)^2\,
(-\vec{q}^{\,2}) \left( \frac{41}{162} 
- \frac{449 \, \vec{q}^{\,2}}{10800} +\ldots\right)\,.
\end{align}
In the integral representation for ${\bar \omega}^{(2)}$ 
given in Eqs.~(15) and~(16) of Ref.~\cite{Pa1993}, one 
should make the replacement 
$\ln\left( \frac{1+ \delta}{1-\delta}\right) \to
\ln\left( \frac{1+ \delta}{1-\delta}\right) \, 
\ln\left( \frac{1+ \delta}{2}\right)$
in order to correct for a typographical error
in an intermediate step of this calculation.
In the coordinate space, the correction becomes
\begin{equation}
V_{\rm vp} = \left[-{\bar \omega}(\nabla^2)
+ {\bar \omega}(\nabla^2)^2+\ldots \right]V\,.
\end{equation}
The contributions to the 
energy involves the first and second order matrix element
together with relativistic corrections,
\begin{align}
\delta E =& \left\langle V_{\rm vp} \right\rangle + 
\left\langle V_{\rm vp}\,\frac{1}{(E-H)'}\, V_{\rm vp}\right\rangle 
\nonumber\\
& + 2\,\left\langle V_{\rm vp}\,\frac{1}{(E-H)'}\, H_R\right\rangle
\nonumber\\
& + \frac{1}{8}\,\left\langle \nabla^2(V_{\rm vp}) 
+ 2\,\sigma^{ij}\,\nabla^i(V_{\rm vp})\,p^j\right\rangle\,.
\end{align}
The two-loop part of this expression reads
\begin{align}
\label{deltaEiv}
\delta^{(2)} E^{iv} =& 
- \frac{41}{162} \,\left(\frac{\alpha}{\pi} \right)^2\, 
\left< \vec{\nabla}^2 V \right> 
\nonumber\\
& - \frac{953}{16200} \, \left(\frac{\alpha}{\pi} \right)^2\,
\left< \vec{\nabla}^2 V \, \frac{1}{(E-H)'}\, \vec{\nabla}^2 V \right> 
\nonumber\\
& +  \frac{41}{648}\,\left(\frac{\alpha}{\pi} \right)^2\, 
\left< \vec{\nabla}^2 V \, \frac{1}{(E-H)'} \, \vec{p}^4 \right>
\nonumber \\ 
& -\frac{41}{1296} \, \left(\frac{\alpha}{\pi}\right)^2\,
\langle 2\,{\rm i}\,\sigma^{ij}\, p^i\,\vec{\nabla}^2V\,p^j\rangle
\nonumber \\ 
&-\frac{557}{8100} \,\left(\frac{\alpha}{\pi}\right)^2\,
\langle\vec{\nabla}^4 V\rangle
\end{align}
The first term in this result corresponds to the 
$\alpha^2\,(Z\,\alpha)^4$ term in Eq.~(\ref{struc2L}).
The remaining terms give the $B^{iv}_{60}$ coefficient.

\begin{table}[htb]
\begin{minipage}{8.6cm}
\caption{\label{tablediv} Values of the difference
$B^{iv}_{60}(nS) - B^{iv}_{60}(1S)$ for the diagrams 
in subset $iv$.}
\begin{tabular}{r@{\hspace*{0.6cm}}.}
\hline
\hline
\rule[-3mm]{0mm}{8mm}
$n$ & 
\multicolumn{1}{c}{$B^{iv}_{60}(nS) - B^{iv}_{60}(1S)$} \\
\hline
2 &     -0.611\,365 \\
3 &     -0.603\,468 \\
4 &     -0.560\,004 \\
5 &     -0.521\,300 \\
6 &     -0.490\,240 \\
\hline
\hline
\end{tabular}
\end{minipage}
\end{table}

We first notice the complete absence of logarithmic terms
in the result (\ref{deltaEiv}).
All matrix elements in (\ref{deltaEiv}) vanish 
for $D$-states and for states with higher angular momenta,
in the order of $\alpha^2\,(Z\alpha)^6$.
The fine-structure difference
of the nonlogarithmic term for $P$
states is as follows,
\begin{equation}
\label{fsPiv}
B^{iv}_{60}(nP_{3/2}) - B^{iv}_{60}(nP_{1/2}) 
= \frac{41}{162} \, \left(1 - \frac{1}{n^2} \right)\,.
\end{equation}
The $n$-dependence of the nonlogarithmic term 
$B^{iv}_{60}$ is as follows,
\begin{align}
\label{ndepB60iv}
& B^{iv}_{60}(nS) - B^{iv}_{60}(1S)
= - \frac{1817}{2025} - \frac{2194}{2025n} + \frac{1337}{675n^2} 
\nonumber\\
& + \frac{2194}{2025} \,
\left[ \gamma + \Psi(n) - \ln(n) \right]\,.
\end{align}
This completes our investigation of the subset $iv$.

%
% TOTAL RESULT FOR ALL TWO--LOOP DIAGRAMS
%
\section{TOTAL RESULT FOR ALL TWO--LOOP DIAGRAMS}
\label{total}

The two-loop subsets $i$---$iv$ (see Figs.~\ref{fig1}---\ref{fig4})
have been considered in Secs.~\ref{calci}---\ref{calciv}.
We are now in the position
to add the results given in Eqs.~(\ref{deltaEi}), (\ref{deltaEii}), 
(\ref{deltaEiii}) and (\ref{deltaEiv}), and to present a general expression
for the complete two-loop correction to the Lamb shift, including the
vacuum-polarization terms, valid for general hydrogenic bound states
with nonvanishing angular momenta, and for the normalized difference
of $S$ states. This general result reads
\begin{widetext}
\begin{align}
\label{delta2LE}
& \delta^{(2)} E =  \delta^{(2)} E^i + \delta^{(2)} E^{ii} +
\delta^{(2)} E^{iii} + \delta^{(2)} E^{iv} 
\nonumber\\
& = \left( \frac{\alpha}{\pi} \right)^2 \frac{(Z\alpha)^4}{n^3} \,
\left\{ B_{40} + (Z\alpha)^2 \, \left[
B_{62}\ln^2[(Z \alpha)^{-2}] + B_{61}\ln[(Z \alpha)^{-2}] + B_{60} 
\right] \right\}
\nonumber\\
& = \left(\frac{\alpha}{\pi}\right)^2\,
\left[ -\frac{2179}{2592}
+ \frac{9}{4}\,\zeta(2)\,\ln 2
- \frac{5}{9}\,\zeta(2)
- \frac{9}{16}\,\zeta(3)\right]\, \left< \vec{\nabla}^2 V \right>
+ \left[\frac{197}{288}
- \frac{3}{2}\,\zeta(2)\,\ln 2
+ \frac{1}{4}\,\zeta(2)
+ \frac{3}{8}\,\zeta(3)
\right]\,
\left< \sigma^{ij}\,\nabla^i V\,p^j \right>
\nonumber\\
& \quad + \left( \frac{\alpha}{\pi} \right)^2 \frac{(Z\alpha)^6}{n^3} \,
\left\{ b_L + \beta_{4} + \beta_{5} + 
\left[\frac{38}{45} + 
\frac{4}{3} \LZa{} \right]\, N \right\}
\nonumber \\ 
& \quad + \left(\frac{\alpha}{\pi}\right)^2
\left[-\frac{42923}{259200} 
+ \frac{9}{16} \zeta(2) \ln (2)
- \frac{5 \zeta(2)}{36} 
- \frac{9 \zeta(3)}{64} 
+ \frac{19}{135} \LZa
+ \frac{1}{9} \LZasquared \right]
\left< \vec{\nabla}^2 V \frac{1}{(E-H)'} \vec{\nabla}^2 V \right> 
\nonumber \\ 
& \quad 
+ \left(\frac{\alpha}{\pi}\right)^2\,
\left[ \frac{2179}{10368} 
- \frac{9}{16} \,\zeta(2)\,\ln (2)
+ \frac{5}{36} \,\zeta(2)
+ \frac{9}{64}\,\zeta(3) \right]\,
\left< \vec{\nabla}^2 V\,\frac{1}{(E-H)'}  \, \vec{p}^{\,4} \right> 
\nonumber \\ 
& \quad
+ \left(\frac{\alpha}{\pi}\right)^2\,
\left[ -\frac{197}{1152} 
+ \frac{3}{8} \,\zeta(2)\,\ln (2)
- \frac{1}{16} \,\zeta(2)
- \frac{3}{32}\,\zeta(3) \right]\,
\left< \vec{p}^{\,4} \,\frac{1}{(E-H)'} \, \sigma^{ij}\,\nabla^i V\,p^j \right> 
\nonumber \\ 
& \quad 
+ \left(\frac{\alpha}{\pi}\right)^2\,
\left[ \frac{233}{576} 
- \frac{3}{4} \,\zeta(2)\,\ln (2)
+ \frac{1}{8} \,\zeta(2)
+ \frac{3}{16}\,\zeta(3) \right]\,
\left< \sigma^{ij}\,\nabla^i V\,p^j \,\frac{1}{(E-H)'} \, 
\sigma^{ij}\,\nabla^i V\,p^j \right> 
\nonumber \\ 
& \quad + \left(\frac{\alpha}{\pi}\right)^2\,
\left[-\frac{197}{2304} 
+ \frac{3}{16} \,\zeta(2)\,\ln (2)
- \frac{1}{32} \,\zeta(2)
- \frac{3}{64}\,\zeta(3)\right]\,
\left< 
\left\{ \vec{p}^{\,2} , \vec{\nabla}^2 V + 2\,\sigma^{ij}\,\nabla^i V\,p^j\right\}
\right> 
\nonumber \\ 
& \quad + \left(\frac{\alpha}{\pi}\right)^2\,
\left[-\frac{83}{1152}  
+ \frac{17}{8} \zeta(2)\,{\ln(2)}
- \frac{59}{72}\,\zeta(2) 
- \frac{17}{32}\,\zeta(3) \right]\,
\left< \left(\vec{\nabla} V \right)^2\right>
\nonumber \\ 
& \quad + \left(\frac{\alpha}{\pi}\right)^2\,
\left[-\frac{87697}{345600} 
+ \frac{9}{10}\zeta(2)\,\ln (2)
- \frac{2167}{9600}\zeta(2)
- \frac{9}{40}\zeta(3)
+ \frac{19}{270} \LZa
+ \frac{1}{18} \LZasquared \right]\,
\left< \vec{\nabla}^4 V \right>
\nonumber \\ 
& \quad + \left(\frac{\alpha}{\pi}\right)^2\,
\left[-\frac{16841}{207360}  
- \frac{1}{5}\, \zeta(2)\,\ln(2)
+ \frac{223}{2880} \, \zeta(2)
+ \frac{1}{20}\,\zeta(3)
+ \frac{1}{24}\, \LZa
\right]\,
\left< 2\,{\rm i}\,\sigma^{ij}\, p^i\,\vec{\nabla}^2V\,p^j\right>\,.
\end{align}
The third line in the above equation corresponds to the 
lower-order $\alpha^2 (Z\alpha)^4$ contribution ($B_{40}$ coefficient).
We now turn to the evaluation of this expression for $S$ states.
The sum of the contributions in 
Eqs.~(\ref{ndepB60i}), (\ref{ndepB60ii}), (\ref{ndepB60iii}) and 
(\ref{ndepB60iv}) corresponds to the sum
of all the matrix elements in Eq.~(\ref{delta2LE}),
evaluated for the normalized difference of $S$ states.
The logarithmic terms~\cite{Pa2001} have already been verified
for the normalized difference.
The $n$-dependence of the total nonlogarithmic term may be 
expressed as
\begin{equation}
\label{defAnS}
B_{60}(nS) - B_{60}(1S) = b_L(nS) - b_L(1S) + A(n), 
\end{equation}
where $A(n)$ is an additional contribution
beyond the $n$-dependence of the two-loop Bethe logarithm,
defined in analogy to Eq.~(\ref{defRnS}).
The result for $A$ is
\begin{align}
\label{An}
& A(n) =
\left( \frac{38}{45} - \frac43\,\ln(2) \right) \, [N(nS) - N(1S)]  
- \frac{337043}{129600} - \frac{94261}{21600\,n} + \frac{902609}{129600\,n^2} 
+ \left( \frac{4}{3}  
- \frac{16}{9\,n}
+ \frac{4}{9\,n^2} \right) \, \ln^2 (2) 
\\
&
+ \left( -\frac{76}{45} 
+ \frac{304}{135\,n}
- \frac{76}{135\,n^2} \right) \,\ln(2) 
+ \left( - \frac{53}{15}
+ \frac{35}{2\,n}
- \frac{419}{30\,n^2} \right) \, \zeta(2)\,\ln(2) 
+ \left( \frac{28003}{10800} 
- \frac{11}{2\,n} 
+ \frac{31397}{10800\,n^2} \right) \, \zeta(2) 
\nonumber\\
& + \left( \frac{53}{60}  
- \frac{35}{8 n}
+ \frac{419}{120 n^2} \right) \,\zeta(3)  
+ \left( \frac{37793}{10800} 
+ \frac{16}{9}\ln^2(2)
- \frac{304}{135}\ln (2) 
+ 8\zeta(2)\ln(2) 
- \frac{13}{3}\zeta(2) 
- 2\zeta(3)\right) 
\left[ \gamma + \Psi(n) - \ln(n) \right] \,.
\nonumber
\end{align}
\end{widetext}
Numerically, $A(n)$ is found to be much smaller than
$b_L(nS) - b_L(1S)$, as shown in Table~\ref{tabletotal}.
This implies that the numerically most important 
contribution to $B_{60}(nS) - B_{60}(1S)$
is exclusively due to the two-loop Bethe logarithm.
The theoretical uncertainty
of $B_{60}(nS) - B_{60}(1S)$, for higher excited $nS$ states,
is caused entirely by the numerical uncertainty of the
two-loop Bethe logarithm $b_L(nS)$, with 
explicitly data for higher excited states taken from Ref.~\cite{Je2004b60}.

\begin{table}[htb]
\begin{minipage}{8.6cm}
\caption{\label{tabletotal} Total values of the difference $B_{60}(nS) -
  B_{60}(1S)$ coming from all diagrams.}
\begin{tabular}{r@{\hspace{1.0cm}}.@{\hspace{1.0cm}}.}
\hline
\hline
\rule[-3mm]{0mm}{8mm}
$n$ &
\multicolumn{1}{c}{$A(n)$} &
\multicolumn{1}{c}{$B_{60}(nS) - B_{60}(1S)$} \\
\hline
2 &      0.x318~486 &   15.x1(4) \\
3 &      0.x360~079 &   18.x3(7) \\
4 &      0.x368~661 &   20.x0(10) \\
5 &      0.x370~042 &   21.x2(11) \\
6 &      0.x369~462 &   22.x0(11) \\
\hline
\hline
\end{tabular}
\end{minipage}
\end{table}

For the fine-structure difference of $D$ states, 
the total two-loop results is obtained by evaluating 
the general result in Eq.~(\ref{delta2LE}) on $D$ states, or alternatively
by adding just the contributions from subsets $i$ and $ii$
[see Eqs.~(\ref{fsDi}) and (\ref{fsDii})],
because the subsets $iii$ and $iv$ do not contribute
to the $D$ fine structure.
For the $P$-state fine structure, the sum of the results in 
Eqs.~(\ref{fsPi}),~(\ref{fsPii}),~(\ref{fsPiii}),
and~(\ref{fsPiv}) gives the complete result, including the 
nonlogarithmic term $B_{60}$.
It has already been stressed
that in order to determine the absolute value of $B_{60}$ for $P$
and $D$ states, an evaluation of the Bethe logarithm $b_L$ for these
states would be required, and its knowledge is currently 
restricted to $S$ states.

Despite this, we may evaluate general logarithmic terms for $P$
and $D$ states. For $D$ states and states with higher angular 
momenta, a direct evaluation of Eq.~(\ref{delta2LE}) immediately
reveals that the logarithmic terms vanish,
\begin{equation}
B_{62}(nD) = B_{61}(nD) = 0 \,.
\end{equation}
The same holds for any hydrogenic states with orbital angular 
momentum $l \geq 2$. For $P$ states, an evaluation of 
(\ref{delta2LE}) confirm that 
\begin{equation}
B_{62}(nP) = \frac{4}{27} \frac{n^2 - 1}{n^2}.
\end{equation}
Furthermore, the logarithmic terms are 
\begin{align}
B_{61}(nP_{1/2}) =& \frac43\, N(nP) + 
\frac{n^2 - 1}{n^2}
\left(\frac{166}{405} -\frac{8}{27} \, \ln 2 \right)\,,
\\
B_{61}(nP_{3/2}) =& \frac43\, N(nP) + 
\frac{n^2 - 1}{n^2}
\left(\frac{31}{405} -\frac{8}{27} \, \ln 2 \right)\,.
\end{align}
Numerical values for $N(nP)$ can be found in Eq.~(17) of
Ref.~\cite{Je2003jpa}.

%
% SUMMARY
%
\section{SUMMARY}
\label{conclusions}

We have presented a unified approach to the one- and two-loop
electron bound-state self-energy correction in hydrogenlike 
atoms, including terms of order $\alpha\,(Z\,\alpha)^6$
and $\alpha^2\,(Z\,\alpha)^6$, respectively.
We consider states with nonvanishing orbital angular momentum
and the normalized difference of $S$ states.
The general analytic structure of the one- and two-loop
corrections is given in Eqs.~(\ref{struc1L}) and (\ref{struc2L}),
respectively. The general result for the one-loop 
correction is given in Eq.~(\ref{delta1LE}).
We evaluate our formulas for specific families of hydrogenic
states in Secs.~\ref{calc1LnD},~\ref{calc1LnP}, and~\ref{calc1LnS}
(one-loop case).
All one-loop results are in agreement with those
previously reported in the literature.
In addition, we obtain results for the nonlogarithmic terms
($A_{60}$ coefficients), for higher excited $S$ states,
as listed in Tab.~\ref{table1}. 

For clarity, we separate the two-loop calculation
into four different subsets $i$, $ii$, $iii$ and $iv$ 
consisting of separately 
gauge-invariant diagrams (see Secs.~\ref{calci}---\ref{calciv}
and Figs.~\ref{fig1}---\ref{fig4}).
A general formula for the ``pure'' two-loop self-energy
diagrams is presented in Eq.~(\ref{deltaEi}).
The corresponding expression, for the 
self-energy vacuum-polarization diagram in Fig.~\ref{fig2}, can be 
found in Eq.~(\ref{deltaEii}). For the subsets 
$iii$ and $iv$, we present general expressions 
in Eqs.~\ref{deltaEiii} and~\ref{deltaEiv}.
For the total sum of the two-loop effects,
a summary is provided in Sec.~\ref{total}.

The two-loop 
fine-structure difference for $P$ states for the subset $i$ as given in 
Eq.~(\ref{fsPi}) is in agreement with previous 
results~\cite{JePa2002,JeKePa2002}.
This constitutes an important cross-check of the method 
used in the current investigation, which is 
based on dimensional regularization, and on effective 
operators for the contributions stemming from hard
virtual photons. The results given in 
Eqs.~(\ref{fsPii}),~(\ref{fsPiii}),
and~(\ref{fsPiv}) complete the fine-structure
difference of $P$ states in the order
$\alpha^2\,(Z\alpha)^6$.

The central result of the current investigation, however, is the 
complete $n$-dependence of all two-loop logarithmic and nonlogarithmic 
contributions to the Lamb shift of $S$ states 
up to the order $\alpha^2 \, (Z\,\alpha)^6$. 
In this regard, our study follows a number of previous
investigations on related 
subjects (see Refs.~\cite{Pa2001,Ka1994,Ka1995,Ka1996b}),
where the logarithmic terms were primarily investigated,
but the nonlogarithmic term was left unevaluated.
The $n$-dependence of all logarithmic terms for $S$ states
[corresponding to the $B_{62}$ and $B_{61}$ coefficients
in Eq.~(\ref{struc2L})] is recovered in full agreement
with the literature.
For the $B_{61}$ coefficient, we refer to Eqs.~(\ref{ndepB61i})
and~(\ref{ndepB61iii}). Moreover we obtain 
in Appendix~\ref{addB61}  an additional logarithmic contribution $B_{61}(1S)$
to the ground 1S state, which was omitted in the former work \cite{Pa2001}.  

Partial results for the $n$-dependence of the 
nonlogarithmic term $B_{60}(nS)$ 
are given in 
Eqs.~(\ref{ndepB60i}),~(\ref{ndepB60i}),~(\ref{ndepB60iii}) 
and~(\ref{ndepB60iv}). A summary including all 
two-loop subsets is provided in 
Eqs.~(\ref{delta2LE}), (\ref{defAnS}) and~(\ref{An}).
Our results lead to predictions for the $S$-state
normalized difference with an accuracy of the order of $100$ Hz
(see Ref.~\cite{CzJePa2005prl} and Appendix~\ref{evalLamb}).
We find that the largest contribution to the $n$-dependence of
$B_{60}$ stems from the two-loop Bethe logarithm $b_L$,
but the remaining contributions in 
Eqs.~(\ref{ndepB60ii}),~(\ref{ndepB60iii}) and.~(\ref{ndepB60iv})
are essential for obtaining complete predictions
(see also Tables~\ref{tabletotal} and~\ref{tableLamb}).

%
% Acknowledgments
%
\section*{Acknowledgments}

We wish to thank Roberto Bonciani for the collaboration
at an early stage of the project.
This work was supported by EU grant No.~HPRI-CT-2001-50034.
A.C.~acknowledges support by Natural Science and Engineering Research Canada.
U.D.J.~acknowledges support from DFG (Heisenberg program)
under contract JE285-1.

\appendix

\section{Electromagnetic form factors}
\label{appa}

We consider the form factors defined by
\begin{equation}
\gamma_\mu \to \Gamma_\mu = 
F_1(q^2)\gamma_\mu + \frac{i}{2m} F_2(q^2) 
\left( {i\over 2} \right)[\qsla, \gamma_\mu]\,,
\label{A1}
\end{equation}
where $q$ is the outgoing photon momentum. The form factors are expanded in
$\alpha$ up to the second order,
\begin{subequations}
\label{ff2}
\begin{eqnarray}
F_1(q^2) &=& 1 +  F_1^{(1)}(q^2)+ F_1^{(2)}(q^2)\,,\\
F_2(q^2) &=&  F_2^{(1)}(q^2)+ F_2^{(2)}(q^2),
\end{eqnarray}
\end{subequations}
where the superscript corresponds to the loop order,
i.e.~to the power of $\alpha$. They have recently been calculated
analytically by Bonciani, Mastrolia and Remiddi in \cite{bmr}.
The results for the form factors expanded into powers of $q^2$ up to
$q^4$ read (in $D=4-2\varepsilon$):
\begin{widetext}
\begin{subequations}
\label{ff}
\begin{eqnarray}
F_1^{(1)}(q^2) &=& \frac{\alpha}{\pi}\,\biggl[q^2 \left( - {1 \over 8} 
- {1 \over 6\varepsilon} - {1 \over 2} \varepsilon \right)
+ q^4 \left( - {11 \over 240} 
- {1 \over 40\varepsilon} 
- {5 \over 48} \varepsilon \right)\biggr]\,,\\
F_2^{(1)}(q^2) &=&  \frac{\alpha}{\pi}\,\biggl[
{1 \over 2} + 2 \varepsilon
+ q^2 \left( {1 \over 12} + {5 \over 12} \varepsilon  \right)
+ q^4 \left( {1 \over 60} + {11 \over 120} \varepsilon  \right)\biggr]\,,\\
F_1^{(2)}(q^2) &=& \Bigl(\frac{\alpha}{\pi}\Bigr)^2\,\biggl\{
q^2 \left[ 
\left( -\frac{1099}{1296} + \frac{77}{144} \zeta(2)\right)_{\rm vp}
- {47 \over 576} 
+ 3 \, \zeta(2) \, \ln 2 
- {175 \over 144} \zeta(2)
- {3 \over 4}\zeta(3) 
\right]
\\
&& + q^4 \left[  \left( -\frac{491}{1440}  + \frac{5}{24} \zeta(2)\right)_{\rm vp}
+ {1721 \over 12960} 
+ {1 \over 72 \, \varepsilon^2}
+ {1 \over 48 \, \varepsilon}
+ {11 \over 10} \, \zeta(2) \, \ln 2 
- {14731 \over 28800} \, \zeta(2) 
- {11 \over 40} \zeta(3) \right]\biggr\} \,,
\nonumber\\
F_2^{(2)}(q^2) &=&  \Bigl(\frac{\alpha}{\pi}\Bigr)^2\,\biggl\{
\left( \frac{119}{36} - 2 \zeta(2)\right)_{\rm vp}
- {31 \over 16} 
- 3 \, \zeta(2) \, \ln 2 
+ {5 \over 2} \, \zeta(2) 
+ {3 \over 4} \, \zeta(3)
\nonumber\\ 
&& + q^2 \left[ \left( \frac{311}{216}  - \frac{7}{8} \zeta(2)\right)_{\rm vp}
-{77 \over 80} 
- {1 \over 12\,\varepsilon} 
- {23 \over 10} \, \zeta(2) \, \ln 2
+ {61 \over 40} \, \zeta(2) 
+ {23 \over 40} \, \zeta(3) 
\right]
\nonumber\\
&& + q^4 \left[ \left( \frac{533}{1080} - \frac{3}{10} \zeta(2)\right)_{\rm vp}
-{1637 \over 5040} 
- {19 \over 720\,\varepsilon}
- {15 \over 14} \, \zeta(2) \, \ln 2 
+ {689 \over 1050} \, \zeta(2) 
+ {15 \over 56} \, \zeta(3) \right]\biggr\}\,.
\end{eqnarray}
\end{subequations}
\end{widetext}
The subscript VP denotes the contribution to the two-loop
form factors which involves a closed fermion loop (see Fig.~\ref{fig2}).
%
% Low-energy limit of the scattering amplitude
%
\section{Low-energy limit of the scattering amplitude}
\label{appb}

\begin{figure}[htb]
\centerline{\includegraphics[width=1.0\linewidth]{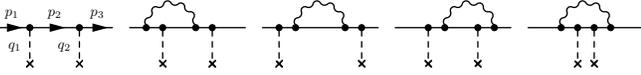}}
\caption{\label{fig5}
Tree and one-loop diagrams with two Coulomb exchanges.}
\end{figure}

In the leading order, the electron self-energy can be incorporated
by electromagnetic form factors $F_1$ and $F_2$, and more precisely by
the leading terms of its low momentum expansion. In the higher order,
namely $\alpha\,(Z\,\alpha)^6$, single vertex form factors $F_i$ are not 
sufficient, and the additional term is the  
low-energy limit of the spin-independent part
of the scattering amplitude with two $\gamma^0$ vertices,
(see Fig.~\ref{fig5}), with the
form factor contributions subtracted. This term, has not yet been 
considered in the literature. A detailed derivation is postponed to a
separate paper; here we present only a brief derivation.

To construct the projection operators for a spin independent part
of the scattering amplitude, let us consider
the matrix element of an arbitrary operator $\hat Q$, namely
$\bar u(p',s')\,\hat Q\,u(p,s)$,
where $u(p,s)$ is a positive solution of the free Dirac equation,
normalized according to $\bar u \,u=1$. We transform this matrix element 
to the more convenient form
%prl.tex
\begin{equation}
\bar u(p',s')\,\hat Q\,u(p,s) = {\rm Tr}[\hat Q\,u(p,s)\,\bar u(p',s')]\,.
\end{equation}
Because we aim to calculate only the low energy limit, we can use
an approximate form of $u(p,s)$,
\begin{eqnarray}
u(p,s) \approx \left(
\begin{array}{l}
{\displaystyle \chi_s}\\[2ex]
{\displaystyle \half\,(\vec\sigma\cdot\vec p)\,\chi_s}
\end{array}
\right)\,,
\end{eqnarray}
where $\chi_s$ is a spinor. Using
\begin{equation}
\sum_s\,\chi_s \, \chi^+_s = I\,,
\end{equation}
where $I$ is the $2 \times 2$ unit matrix,
the spin-averaged projection operator becomes
\begin{eqnarray}
\sum_s u(p,s)\,\bar u(p',s) &\approx& \left(
\begin{array}{lr}
{\displaystyle I}&\quad 
{\displaystyle -\frac{\vec\sigma\cdot\vec p'}{2}}\\[4ex]
{\displaystyle \frac{\vec\sigma\cdot\vec p}{2}}&\quad
{\displaystyle -\frac{\vec\sigma\cdot\vec p}{2}\,
\frac{\vec\sigma\cdot\vec p'}{2}}
\end{array}\right)
\nonumber\\
& \approx &
\frac{\not\!p+1}{2}\,\frac{\not\!p'+1}{2}\,.
\end{eqnarray}
The spin-averaged matrix element of an arbitrary
operator $\hat Q$ can now be expressed as
\begin{equation}
\langle\hat Q\rangle = \frac{1}{8}\,{\rm Tr}\Bigl[
(\not\!p'+1)\,\hat Q\,(\not\!p+1)
\Bigr]\,.
\end{equation}

We can now turn to the scattering amplitude $T$.
The expression corresponding to the tree diagram of Fig.~\ref{fig5} is
\begin{equation}
\label{Ttree}
T^{(0)} =
\frac{1}{8}\,{\rm Tr}\left[(\not\!p_1+1)\,\gamma_0\,\frac{1}{\not\!p_2-1}\,
\gamma_0\,(\not\!p_3+1)\right]\,,
\end{equation}
and this expression defines our normalization.
The presence of $\gamma_0$ in Eq.~(\ref{Ttree}) results from the fact,
that we consider the scattering by the Coulomb potential
\begin{equation}
\label{VC}
e \gamma^\mu A_\mu = \gamma_0 \, V  = 
-\gamma_0 \, \frac{Z\,e^2}{\vec{q}^{\,2}}\,.
\end{equation}
The momenta $p_1$ and $p_3$ are on mass shell
($ p_1^2 = 1$, $p_3^2 = 1$).
Let us define the exchange momenta according to
\begin{equation}
q_1 = p_1-p_2\,, \qquad
q_2 = p_2-p_3\,,
\end{equation}
and the static momentum $t$, such that $t = (1,\vec 0)$ and $t^2 = 1$.
Because we consider the scattering of a static potential,
the exchange momenta are spatial,
\begin{equation}
q_1^\mu  t_\mu = q_2^\mu  t_\mu = 0 \,.
\end{equation}
The one- and two-loop radiative corrections, $T^{(1)}$ and $T^{(2)}$,
are obtained using standard rules of quantum electrodynamics.
However, we additionally subtract from these amplitudes 
the corresponding form factor contribution.
This subtraction is carried out using
the tree diagram with the vertex $\gamma_0$
replaced by $\Gamma_0$,
\begin{equation}
\frac{1}{8}\,{\rm Tr}\left[(\not\!p_1+1)\,\Gamma_0(q_1)
\,\frac{1}{\not\!p_2-1}\,
\Gamma_0(q_2)\,(\not\!p_3+1)\right]\,.
\end{equation}
The vertex function $\Gamma^\mu$ is defined in Eq.~(\ref{A1}).
In the one-loop order, the subtraction permits the 
approximation $\Gamma^\mu \approx 1$ for one of the vertices, 
with a form-factor correction at the other, and a second 
term where the approximations at the vertices are interchanged.
For the two-loop case, it is understood that the subtraction includes
only $(\alpha/\pi)^2$ terms, so there are a total of three terms,
one with both vertices modified
by one loop corrections, and two others where only one vertex 
receives a two-loop correction.
After the form-factor subtractions and small momenta expansion, 
the scattering amplitude takes a simple form
\begin{equation}
T^{(i)} = q_1 \cdot q_2\,\chi^{(i)}\,,
\end{equation}
where the superscript denotes the loop order.
The coefficients $\chi$ have been calculated with the help of
the symbolic program {\sl FORM} \cite{form} and read
\begin{subequations}
\label{eta1plus2}
\begin{eqnarray}
\label{eta1}
\chi^{(1)} &=& \left(\frac{\alpha}{\pi}\right)\,
\left( \frac{1}{6} - \frac{1}{3\,\varepsilon}\right)\,,\\
\label{eta2}
\chi^{(2)} &=&
\left(\frac{\alpha}{\pi}\right)^2\,\left[
- \frac{79}{288}  
+ \frac{5}{2}\,\zeta(2)\,\ln(2)
- \frac{127}{144}\,\zeta(2) 
\right.
\nonumber\\[2ex]
& & \left.  - \frac{5}{8}\,\zeta(3)
+ \left(-\frac{391}{648}  + \frac{205}{576}\,\zeta(2) \right)_{\rm vp}
\right]\,.
\end{eqnarray}
\end{subequations}
where the subscript $\rm vp$ denotes the contribution from the
diagram in Fig.~\ref{fig2}. 
Using the relation $q_1 \cdot q_2 = - \vec{q}_1 \cdot \vec{q}_2$,
and including the factors given by the Coulomb potential, one obtains
the effective interaction Hamiltonian
\begin{align}
\delta H =& -(Z\,e^2)^2 \,
\frac{\vec{q}_1 \cdot \vec{q}_2}{\vec{q}_1^{\,2} \, \vec{q}_2^{\,2}}\,\chi
\nonumber\\[2ex]
\rightarrow&
-(-{\rm i} \vec{\nabla} V)^2\,\chi = (\vec{\nabla} V)^2\,\chi = 
e^2\,{\vec E}^2\,\chi \,.
\end{align}
where by $\rightarrow$ we denote the transition to the coordinate 
space by the corresponding Fourier transform.

%
% ADDITIONAL LOGARITHMIC CONTRIBUTION TO THE GROUND STATE LAMB SHIFT
%
\section{ADDITIONAL LOGARITHMIC CONTRIBUTION TO THE GROUND STATE LAMB SHIFT}
\label{addB61}

The two-loop logarithmic contribution to the Lamb shift has been considered
by one of us (K.P.) in \cite{Pa2001}. The obtained results for $B_{61}$
coefficient of the ground state was
\begin{eqnarray}
B_{61}^{\rm old} &=& \frac{39751}{10800} + \frac{4}{3}\,N(1S) + 
\frac{55\,\pi^2}{27} - \frac{616\,\ln(2)}{135} \nonumber \\ &&
+ \frac{3\,\pi^2\,\ln(2)}{4} 
+ \frac{40\,\ln^2(2)}{9} - 
\frac{9\,\zeta(3)}{8} \nonumber \\ &=&  50.309\,654\,.
\end{eqnarray}
After careful reanalysis of the performed calculations we found that
there is an additional logarithmic contribution,  which can be associated
to the $e^2\,\vec E^2 = (\vec{\nabla} V)^2$ term in the effective Hamiltonian 
in Eqs. (\ref{trafo2},\ref{deltaH}) 
\begin{equation}
\delta H =\left[\frac{F^{\rm (2)}_2(0)+\bigl(F_2^{(1)}\bigr)^2}{8}+\chi^{(2)}\right]
\,(\vec{\nabla} V)^2
\end{equation}
Although the coefficient is finite, the matrix element of $(\vec{\nabla} V)^2$
yields the logarithm
\begin{equation}
\langle(\vec{\nabla} V)^2\rangle \simeq -4\,\frac{(Z\,\alpha)^6}{n^3}\,\ln[(Z\,\alpha)^{-2}]
\end{equation}
The additional contribution to $B_{61}$ is therefore
\begin{eqnarray}
\delta B_{61} &=& -4\,\left[\frac{F^{\rm
      (2)}_2(0)+\bigl(F_2^{(1)}\bigr)^2}{8}+\chi^{(2)}\right]
\nonumber \\
 &=&\frac{559}{288} + \frac{41}{18}\,\zeta(2) -
   \frac{17}{2}\,\ln (2)\,\zeta(2) + \frac{17}{8}\,\zeta(3)\nonumber \\ &&
+\left(\frac{493}{648} - \frac{61}{144}\,\zeta(2)\right)_{\rm vp} = -1.385\,414\,.
\end{eqnarray}
Again, the subscript $\rm vp$ denotes the contribution 
from the subset $ii$ of two-loop diagrams (Fig.~\ref{fig2}).
The new value for the logarithmic contribution including 
the vacuum polarization is
\begin{eqnarray}
B_{61} &=& B^{\rm old}_{61} + \delta B_{61} = 
\frac{413581}{64800} + \frac{4}{3}\,N(1S) + \frac{2027}{864}\,{\pi }^2 
\nonumber \\ &&
-\frac{616}{135}\,\ln (2) - \frac{2}{3}\,{\pi }^2\,\ln (2) +
   \frac{40}{9}\,{\ln^2 (2)} + {\zeta}(3) 
\nonumber \\ &=& 48.958\,590\,.
\end{eqnarray}
Since this additional contribution is numerically small, it does not
explain the discrepancy with the direct numerical
calculation by Yerokhin {\em et al.} in Ref.~\cite{YeInSh2005}, although
the difference is now slightly smaller. We postpone further conclusions
until the evaluation of the constant term $B_{60}$ is completed.

%
% EVALUATION OF THE LAMB--SHIFT DIFFERENCE
%
\section{EVALUATION OF THE LAMB--SHIFT DIFFERENCE}
\label{evalLamb}

We denote the Lamb shift of an $nS$ states by $\Delta E(nS)$
and use the definition in Eq.~(67) of Ref.~\cite{JePa1996}.
We focus on the evaluation of the normalized difference for 
$S$ states, which we denote as 
\begin{equation}
\label{defDelta}
\Delta_n \equiv n^3 \,\Delta E(nS) - \Delta E(1S)\,.
\end{equation}
Important contributions to the Lamb shift as used for the 
data in Table~\ref{tableLamb}, can be found in  
Tables 1---10 of Ref.~\cite{EiGrSh2001}.
The new results derived in this article for the 
nonlogarithmic two-loop term $B_{60}(nS) - B_{60}(1S)$ 
can now be used for an improvement of the accuracy of the 
theoretical predictions as listed in Table~\ref{tableLamb}.

Extrapolations of the two-loop Bethe logarithms $b_L(nS)$, and of the 
$A_{60}$ coefficients in Table~\ref{table1}, to higher principal quantum 
numbers, are performed by assuming a functional form
of the type $a + b/n + c/n^2$ for the correction, with $a$, $b$ and 
$c$ as constant coefficients. This functional form has recently
been shown to be applicable to a variety of quantum 
electrodynamic corrections for bound states,
see e.g.~Refs.~\cite{Je2003jpa,JeEtAl2003}.
The same functional forms are used to extrapolate
the difference $G_{\rm SE}(\alpha) - A_{60}$,
as a function of $n$, to higher 
principal quantum numbers [numerical results of the 
nonperturbative self-energy remainder $G_{\rm SE}(\alpha)$
can be found in Refs.~\cite{JeMoSo2001pra,JeMo2004pra}].

\begin{table}[htb]
\begin{minipage}{8.6cm}
\caption{\label{tableLamb} Theoretical values of the normalized
Lamb-shift difference $\Delta_n$ defined in Eq.~(\ref{defDelta}),
using results obtained here [see Eq.~(\ref{An})].
Units are kHz.}
\begin{tabular}{r@{\hspace*{0.5cm}}.r@{\hspace*{0.5cm}}.}
\hline
\hline
$n$ & \multicolumn{1}{c}{$\Delta_n$} &
$n$ & \multicolumn{1}{c}{$\Delta_n$} \\
\hline
2 &    187225.x70(5) & 17 &    281845.x77(11) \\
3 &    235070.x90(7) & 18 &    282049.x05(11) \\
4 &    254419.x32(8) & 19 &    282221.x81(11) \\
5 &    264154.x03(9) & 20 &    282369.x85(11) \\
6 &    269738.x49(9) & 21 &    282497.x67(11) \\
7 &    273237.x83(9) & 22 &    282608.x78(11) \\
8 &    275574.x90(10) & 23 &   282705.x98(11) \\
9 &    277212.x89(10) & 24 &   282791.x50(11) \\
10 &   278405.x21(10) & 25 &   282867.x11(11) \\
11 &   279300.x01(10) & 26 &   282934.x29(11) \\
12 &   279988.x60(10) & 27 &   282994.x18(11) \\
13 &   280529.x77(10) & 28 &   283048.x01(11) \\
14 &   280962.x77(10) & 29 &   283096.x35(11) \\
15 &   281314.x61(10) & 30 &   283140.x01(11) \\
16 &   281604.x34(11) & 31 &   283179.x54(11) \\
\hline
\hline
\end{tabular}
\end{minipage}
\end{table}

The principal theoretical uncertainty with regard to the
normalized difference $\Delta_n$ currently originates
from the unknown $n$-dependence of the two-loop coefficient 
$B_{71}(nS)$. An estimate for this correction may be obtained as follows.
We first map the one-loop coefficient $A_{50}$ onto an
effective Dirac delta potential $V_{50}$, with
\begin{equation}
V_{50} = \frac{\alpha}{\pi} (Z\alpha)^2 \,
\left[ \frac{427}{384} - \frac12\,\ln 2 \right]\,
\vec{\nabla}^2 V\,,
\end{equation}
Of course, $\vec{\nabla}^2 V = 4 \pi \delta^3(r)$,
and we may use this potential as an ``input'' for 
evaluation the additional 
one-loop correction to the Bethe logarithm
generated by the local potential.
This leads to a correction of order $\alpha^2\,(Z\alpha)^7$,
with logarithmic terms. The leading double-logarithmic 
term (corresponding to a $B_{72}$-coefficient) is $n$-independent.
The well-known 
$n$-dependence of the single logarithm,
which gives rise to a $B_{71}$-coefficient,
may be found e.g.~in Eq.~(20) of Ref.~\cite{Pa2001}.
The calculation leads to the estimate
\begin{align}
& B_{71}(nS) - B_{71}(1S) \approx
\pi  \, \left( \frac{427}{36} - \frac{16}{3} \ln(2) \right) 
\nonumber\\
& \times \left[ \frac{3}{4} - \frac{1}{n} + \frac{1}{4 n^2} +
\gamma + \Psi(n) - \ln(n) \right]
\end{align}
for the $nS$-$1S$ difference of the logarithmic term.
As an uncertainty estimate for $B_{71}(nS) - B_{71}(1S)$,
we take half the value of the above expression.


\begin{thebibliography}{10}

\bibitem{Mo1974a}
P.~J. Mohr, Ann. Phys. (N.Y.) {\bf 88},  26  (1974).

\bibitem{Mo1974b}
P.~J. Mohr, Ann. Phys. (N.Y.) {\bf 88},  52  (1974).

\bibitem{YeInSh2005}
V.~A. Yerokhin, P. Indelicato, and V.~M. Shabaev, Phys. Rev. A {\bf 71},
  R040101  (2005).

\bibitem{Pa2001}
K. Pachucki, Phys. Rev. A {\bf 63},  042503  (2001).

\bibitem{PiSo1998}
A. Pineda and J. Soto, Phys. Lett. {\bf B420}, 391 (1998).

\bibitem{ItZu1980}
C. Itzykson and J.~B. Zuber, {\em Quantum Field Theory} (McGraw-Hill, New York,
  NY, 1980).

\bibitem{Pa2004}
K. Pachucki, Phys. Rev. A {\bf 69},  052502  (2004).

\bibitem{Pa1993}
K. Pachucki, Ann. Phys. (N.Y.) {\bf 226},  1  (1993).

\bibitem{JePa1996}
U. Jentschura and K. Pachucki, Phys. Rev. A {\bf 54},  1853  (1996).

\bibitem{JeSoMo1997}
U.~D. Jentschura, G. Soff, and P.~J. Mohr, Phys. Rev. A {\bf 56},  1739
  (1997).

\bibitem{EiGrSh2001}
M.~I. Eides, H. Grotch, and V.~A. Shelyuto, Phys. Rep. {\bf 342},  63  (2001).

\bibitem{MoTa2005}
P.~J. Mohr and B.~N. Taylor, Rev. Mod. Phys. {\bf 77},  1  (2005).

\bibitem{JeEtAl2003}
U.~D. Jentschura, E.-O. Le~Bigot, P.~J. Mohr, P. Indelicato, and G. Soff, Phys.
  Rev. Lett. {\bf 90},  163001  (2003).

\bibitem{JeMoSo1999}
U.~D. Jentschura, P.~J. Mohr, and G. Soff, Phys. Rev. Lett. {\bf 82},  53
  (1999).

\bibitem{JeMo2004pra}
U.~D. Jentschura and P.~J. Mohr, Phys. Rev. A {\bf 69},  064103  (2004).

\bibitem{PaJe2003}
K. Pachucki and U.~D. Jentschura, Phys. Rev. Lett. {\bf 91},  113005  (2003).

\bibitem{Je2004b60}
U.~D. Jentschura, Phys. Rev. A {\bf 70},  052108  (2004).

\bibitem{Je2003jpa}
U.~D. Jentschura, J. Phys. A {\bf 36},  L229  (2003).

\bibitem{Ka1996}
S.~G. Karshenboim, J. Phys. B {\bf 29},  L29  (1996).

\bibitem{JeNa2002}
U.~D. Jentschura and I. Nandori, Phys. Rev. A {\bf 66},  022114  (2002).

\bibitem{JePa2002}
U.~D. Jentschura and K. Pachucki, J. Phys. A {\bf 35},  1927  (2002).

\bibitem{JeKePa2002}
U.~D. Jentschura, C.~H. Keitel, and K. Pachucki, Can. J. Phys. {\bf 80},  1213
  (2002).

\bibitem{Je2003plb}
U.~D. Jentschura, Phys. Lett. B {\bf 564},  225  (2003).

\bibitem{Sc1970}
J. Schwinger, {\em Particles, Sources and Fields} (Addison-Wesley, Reading, MA,
  1970).

\bibitem{BaRe1973}
R. Barbieri and E. Remiddi, Nuovo Cim. A {\bf 13},  99  (1973).

\bibitem{Ka1994}
S.~G. Karshenboim, Zh. \'{E}ksp. Teor. Fiz. {\bf 106},  414  (1994), [JETP {\bf
  79}, 230 (1994)].

\bibitem{Ka1995}
S.~G. Karshenboim, Yad. Fiz. {\bf 56},  707  (1995), [Phys. At. Nucl. {\bf 58},
  649 (1995)].

\bibitem{Ka1996b}
S.~G. Karshenboim, JETP {\bf 82},  403  (1996), [ZhETF {\bf 109}, 752 (1996)].

\bibitem{CzJePa2005prl}
A. Czarnecki, U.~D. Jentschura, and K. Pachucki, 
  Phys. Rev. Lett. {\bf 95}, 180404 (2005).

\bibitem{JeMoSo2001pra}
U.~D. Jentschura, P.~J. Mohr, and G. Soff, Phys. Rev. A {\bf 63},  042512
  (2001).

\bibitem{bmr}
R. Bonciani, P. Mastrolia and E.Remiddi,
\newblock{Nucl.Phys. {\bf B661}, 289 (2003); 
          Erratum-ibid. {\bf B702} 359 (2004).}

\bibitem{form}
J.A.M. Vermaseren, math-ph/0010025.

\end{thebibliography}
\end{document}